\documentclass[reprint,prb,aps,amsmath,amssymb,superscriptaddress]{revtex4-2}

\usepackage[T1]{fontenc}
\usepackage[utf8]{inputenc}
\usepackage[english]{babel}
\usepackage{bm}
\usepackage{graphicx}
\usepackage{units}
\usepackage{hyperref}
\usepackage{orcidlink}
\hypersetup{
  colorlinks=true,
  citecolor=blue,
  urlcolor=blue,
  linkcolor=red,
  breaklinks=true
}
\begin{document}

\title{Mode-Selective Cloaking and Phase-Matching Cavity Resonances in Bilayer Graphene Transport}
\author{Dan-Na Liu
\orcidlink{0009-0009-9351-2104}
}
\affiliation{School of Physics, Xidian University, Xi$'$an, Shaanxi 710071, China}
\affiliation{IMDEA Nanoscience, C/ Faraday 9, 28049 Madrid, Spain}
\author{Jun Zheng
\orcidlink{0000-0001-8426-473X}
}
\affiliation{College of Physics and Technology, Bohai University, Jinzhou, Liaoning 121013, China}
\author{Pierre A. Pantale\'on
\orcidlink{0000-0003-1709-7868}
}
\email{pierre.pantaleon@imdea.org}
\affiliation{IMDEA Nanoscience, C/ Faraday 9, 28049 Madrid, Spain}

\begin{abstract}
We study ballistic electron transport through electrostatic barriers in AB-stacked bilayer graphene within a full four-band framework. A mode-resolved analysis reveals how propagating and evanescent channels couple across electrostatic interfaces and how channel selectivity governs transport at normal incidence. We show that perfect transmission can occur at discrete energies due to phase matching of a single internal mode within an individual barrier, without activating the decoupled channels. This effect is interpreted as a phase-matching cavity, namely an effective cavity formed by internal phase coherence inside the barrier, which yields perfect transmission at discrete energies without true bound states and without opening additional transport channels. For single- and double-barrier geometries, we derive compact analytical expressions for the transmission and identify the corresponding resonance conditions. Extending the analysis to multibarrier structures using a transfer-matrix approach, we demonstrate how perfect resonances driven by internal phase matching coexist with Fabry-Perot-like resonances arising from inter-barrier interference. Our results provide a unified, channel-resolved description of tunneling suppression and resonance-assisted transport in bilayer graphene barrier systems.
\end{abstract}

\maketitle

\section{Introduction}

Bilayer graphene (BG) junctions have become a versatile platform for exploring quantum transport phenomena in low-dimensional systems. Recent experiments using electrostatically defined tunnel junctions have demonstrated resonant tunneling and negative differential resistance in vertical double-bilayer graphene heterostructures, underscoring the sensitivity of BG transport to interband coupling and evanescent states~\cite{Fallahazad2014Gate,Burg2018Strongly}. In lateral device geometries, gate-defined cavities in gapped BG have revealed ballistic Fabry-Pérot interference arising from phase-coherent transport~\cite{Varlet2014Fabry}. More recently, Corbino-geometry measurements have reported conductance signatures consistent with angularly selective tunneling in BG, providing clean access to bulk transport properties without edge contributions~\cite{Elahi2024Direct}. Together, these experiments establish BG junctions as a controllable setting in which transport is governed by band alignment, evanescent modes, and the multiband character of the electronic spectrum~\cite{Gayduchenko2021Tunnel}. Ballistic transport across electrostatic barriers offers a particularly transparent framework for probing quantum interference, chirality, and mode selectivity in graphene-based systems~\cite{Beenaker2008Colloquium}. In graphene, chiral quasiparticles give rise to unconventional tunneling phenomena rooted in relativistic quantum mechanics~\cite{Klein1929Die,Katsnelson2006Chiral,Pereira2010Klein}. While monolayer graphene hosts massless Dirac fermions, AB-stacked BG supports massive chiral quasiparticles and an electrically tunable band gap~\cite{mccann2006landau,castro2007biased,Varlet2015Band}. These features lead to transport behavior that differs qualitatively from both monolayer graphene and conventional Schrödinger electrons, including strong angular dependence and mode-specific transmission characteristics~\cite{Gu2011Chirality,Yamamoto1989Transmission,Wu1991Quantum,Ulloa1990Ballistic}. In particular, whereas monolayer graphene exhibits Klein tunneling with perfect transmission at normal incidence due to pseudospin conservation~\cite{Katsnelson2006Chiral}, bilayer graphene instead displays symmetry-protected cloaking and transmission suppression arising from its four-band structure~\cite{Gu2011Chirality}. A defining property of BG is its full four-band low-energy electronic structure. At a given energy, multiple longitudinal solutions coexist, corresponding to both propagating and evanescent modes~\cite{Nilsson2007Transmission,van2013four}. Transport across electrostatic interfaces is therefore governed by selective coupling between external propagating states and internal modes supported within the barrier region~\cite{Gu2011Chirality,Huang2025Evanescent}. Symmetry constraints play a central role in this coupling. In particular, at normal incidence certain internal solutions become symmetry-decoupled from incident propagating channels, leading to a pronounced suppression of transmission even when states are available inside the barrier. This mode-selective decoupling constitutes the microscopic origin of tunneling suppression in BG and is commonly referred to as the cloaking effect~\cite{Gu2011Chirality}. Although related suppression phenomena have also been discussed under the label of anti-Klein tunneling~\cite{Katsnelson2006Chiral,Varlet2014Fabry,AgrawalGarg2012Reversal,Chen2009Design}, a channel-resolved four-band description is essential for capturing the full transport behavior of BG junctions. Transport through single and multiple electrostatic barriers in BG has been widely studied, revealing tunneling suppression, resonant transmission, and interference effects~\cite{Snyman2007Ballistic,Nilsson2007Transmission,van2013four,barbier2009bilayer,Lee2016Evidence,Chen2009Design}. In multibarrier geometries, Fabry-Pérot-like resonances arise from phase-coherent propagation between successive barriers, in close analogy with semiconductor heterostructures~\cite{Ulloa1990Ballistic,Wu1991Quantum}. At the same time, the existence of internal propagating solutions within individual barriers suggests additional resonance mechanisms that are absent in reduced two-band descriptions~\cite{Gu2011Chirality,Park2011Pi}. While recent experiments report conductance signatures consistent with angular selectivity and tunneling suppression~\cite{Elahi2024Direct}, the measured observable is the total conductance, integrated over all incident angles and transport channels. As a result, the microscopic role of individual propagating and evanescent modes, and their selective coupling or decoupling at electrostatic interfaces, remains implicit. Despite substantial progress, the relationship between mode-selective decoupling and resonant transmission therefore remains incompletely resolved at the microscopic level. While early studies established tunneling suppression at normal incidence~\cite{Katsnelson2006Chiral,Gu2011Chirality}, later works reported resonant transmission features in single- and multibarrier structures~\cite{Lu2015Destruction,Lamas2024Persistent}. It has remained unclear whether such resonances indicate a breakdown of channel-selective decoupling or instead originate entirely from phase coherence within the subset of non-decoupled transport channels. A unified analytical framework capable of separating internal phase-matching mechanisms from inter-barrier interference effects is thus still lacking. In this work, we address these issues by studying ballistic electron transport in AB-stacked bilayer graphene within a full four-band framework and performing a systematic mode-resolved analysis of propagating and evanescent channels in electrostatic barrier geometries. Using a transfer-matrix formalism, we show that phase matching of a single non-decoupled internal mode within an individual barrier can yield perfect transmission at discrete energies, while symmetry-imposed decoupling of cloaked channels remains intact. We interpret this mechanism as a phase-matching cavity, namely an effective cavity formed by internal phase coherence within the barrier without true bound states and without restoring coupling to decoupled channels. This unified treatment clarifies how internal phase matching and interference effects coexist in bilayer graphene junctions and provides a channel-resolved understanding of tunneling suppression and resonance-assisted transport. The paper is organized as follows. In Sec.~II we introduce the theoretical framework and classify the propagating and evanescent modes relevant for ballistic transport across electrostatic barriers in bilayer graphene. In Sec.~III we analyze cloaking and confinement effects at normal incidence and introduce the phase-matching cavity mechanism responsible for perfect resonant transmission. In Sec.~IV we extend the analysis to multibarrier structures and distinguish between perfect resonances arising from internal phase matching and Fabry–Pérot–like resonances generated by inter-barrier interference. Finally, Sec.~V presents a discussion of the results and their broader implications.

\section{Propagating and evanescent modes}
The theoretical framework describing ballistic transport across electrostatic barriers in graphene-based systems is well established~\cite{Snyman2007Ballistic,Nilsson2007Transmission,Huang2025Evanescent,van2013four,barbier2009bilayer,Zhao2011Electronic,Wang2011Robust}. Rather than repeating standard derivations, we summarize here only the elements required to classify transport channels and to establish the notation used throughout this work. Technical details of the transfer-matrix formalism are provided in App.~\ref{App: AMatrix}. For clarity, we first focus on a single-barrier geometry, which captures the essential transport mechanisms, and later extend the discussion to multibarrier structures.
\begin{figure}
  \centering
  \includegraphics[width=1.0\columnwidth]{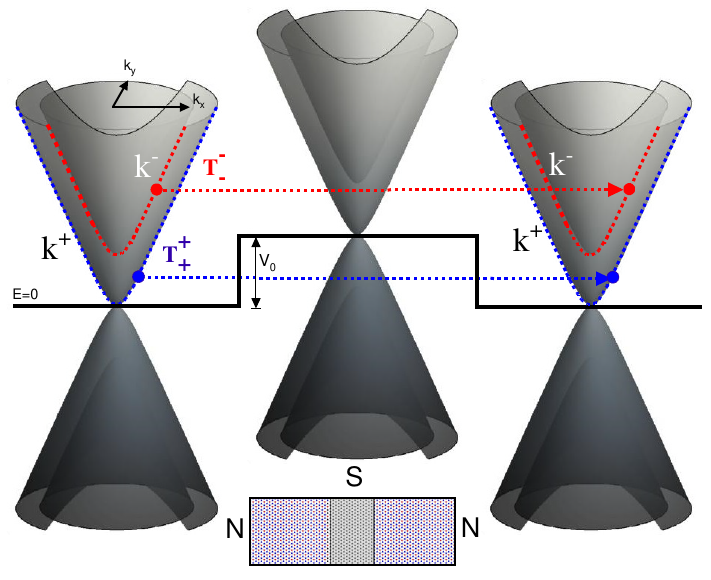}
  \caption{{\bf Quantum transport in BG.} Schematics of the electronic structure of AB-stacked BG in the presence of a single electrostatic barrier. The left and right $N$ regions are unperturbed, while the central $S$ region is subjected to a uniform on-site electrostatic potential $V_{0}$, producing a rigid shift of the energy bands. At normal incidence, transport occurs through two independent channels, $T^+_+$ and $T^-_-$, corresponding to non-scattering processes $k^{\pm}\rightarrow k^{\pm}$, as indicated by the dashed arrows in the figure. Below the bands in the S region we show the schematics of a two-terminal BG device with a single barrier.}
  \label{fig:1}
\end{figure}
We consider ballistic electron propagation in AB-stacked BG in the presence of electrostatic barriers, as schematically shown in Fig.~\ref{fig:1}. Such barriers can be realized experimentally using metallic gates that generate local electrostatic potentials~\cite{Jing2010Quantum,Elahi2024Direct}. The low-energy electronic properties of AB-stacked BG are described using an effective four-band Hamiltonian near the $K$ valley, as described in the App.~\ref{App: AMatrix}. The system is translationally invariant along the transverse $y$ direction, so that the transverse momentum $k_y$ is conserved. Transport therefore reduces to a one-dimensional scattering problem along the longitudinal $x$ direction.

As a representative example, we consider a single-barrier configuration in which the system is divided into three regions: an unperturbed left region ($x<0$), a central electrostatically modified region ($0<x<L$), and an unperturbed right region ($x>L$). The width of the barrier region is denoted by $L$, and the regions are labeled $N$, $S$, and $N$, respectively. Each region is described by the same $4\times4$ BG Hamiltonian described below, with the inclusion of a uniform electrostatic potential $V_0$ in the $S$ region, which rigidly shifts the local band structure. The low-energy electronic properties of AB-stacked BG are described using an effective four-band Hamiltonian near the $K$ valley. To obtain compact analytical expressions, we neglect the trigonal warping terms $\gamma_3$ and $\gamma_4$, which do not affect the normal-incidence results discussed here. In the basis $(A_1,B_1,B_2,A_2)$, the Hamiltonian reads~\cite{mccann2006landau,Mayorov2011Interaction,Kleptsyn2015Chiral}
\begin{equation}
H=
\begin{pmatrix}
V_0 & \hbar v_f \pi & \gamma_1 & 0 \\
\hbar v_f \pi^{+} & V_0 & 0 & 0 \\
\gamma_1 & 0 & V_0 & \hbar v_f \pi^{+} \\
0 & 0 & \hbar v_f \pi & V_0
\end{pmatrix},
\quad
\psi=
\begin{pmatrix}
\psi_{A_1} \\
\psi_{B_1} \\
\psi_{B_2} \\
\psi_{A_2}
\end{pmatrix},
\label{eq: BilHamil}
\end{equation}
where $\pi = k_x + i k_y$ with $\boldsymbol{k}=(k_x,k_y)$ and $v_f~\approx~10^6$~m/s the Fermi velocity of graphene. The parameter $V_0$ denotes a electrostatic potential and $\gamma_1 =0.4$ eV the interlayer hopping parameter. Because $k_y$ is conserved, the wavefunction can be written as $\psi(x,y)=\Phi(x)e^{i k_y y}$, where the four-component spinor $\Phi(x)=[\phi_{A_1}(x),\phi_{B_1}(x),\phi_{B_2}(x),\phi_{A_2}(x)]^{T}$ describes the sublattice amplitudes on the two graphene layers. Solving the eigenvalue equation $H\Phi(x)=E\Phi(x)$ in the unperturbed $N$ regions (see App.~\ref{App: AMatrix} for further details) yields the four-band energy spectrum
\begin{equation}
E=\pm \left( \frac{\gamma_{1}}{2}\pm\sqrt{k^2+\frac{\gamma_{1}^{2}}{4}} \right).
\label{Eq: Epristine}
\end{equation}
Restricting to electron-like states with $E>0$, this dispersion admits two longitudinal solutions of the form $k^{\pm}=\sqrt{E(E\pm\gamma_1)-k_y^2}$. Real values of $k^{\pm}$ correspond to propagating modes, while imaginary values correspond to evanescent modes. At normal incidence ($k_y=0$), both $k^{+}$ and $k^{-}$ are propagating for $E>\gamma_1$, whereas only the $k^{+}$ mode propagates for $E<\gamma_1$.

In the presence of an electrostatic barrier, the four-band structure naturally gives rise to four transport channels: two non-scattering channels, $T^+_+: k^{+}\rightarrow k^{+}$ and $T^-_-: k^{-}\rightarrow k^{-}$, and two scattering channels, $T^+_-: k^{+}\rightarrow k^{-}$ and $T^-_+: k^{-}\rightarrow k^{+}$~\cite{van2013four}. At normal incidence, symmetry constraints suppress mode mixing, so that only the channels $T^+_+$ and $T^-_-$ contribute to transport. We emphasize that at $k_y=0$, this suppression is exact for any scalar electrostatic profile $V(x)$ that preserves layer-exchange symmetry. The transfer matrix factorizes into two independent $2\times2$ sectors, yielding $T^{+}_{-}=T^{-}_{+}=0$ identically. A compact proof is given in App.~\ref{App: Smooth}. In the $N$ regions, only propagating modes are present, whereas in the barrier region $S$ both propagating and evanescent modes may appear and participate in the scattering process~\cite{Ando1991Quantum,Wu1991Quantum,Ulloa1990Ballistic}.

In the following sections, we show how the coexistence of propagating and evanescent internal modes inside the electrostatic barrier, together with their channel-selective coupling to incident states, gives rise to distinct transport regimes. 

\begin{figure*}
  \centering 
  \includegraphics[width=0.80 \textwidth]{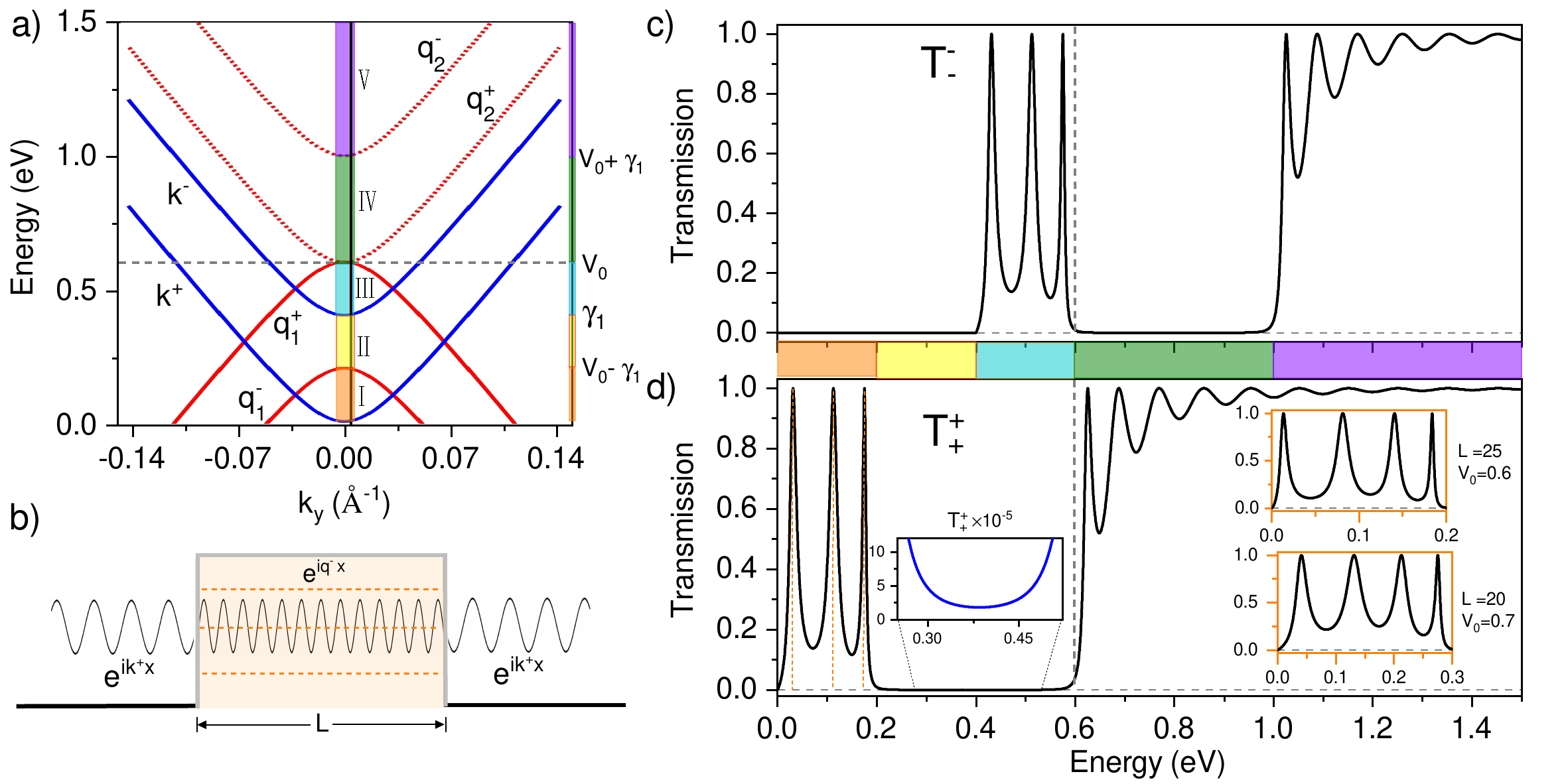}
  \caption{{\bf Transport modes.} Band structure in the $N$ region (blue lines) with modes $k^{\pm}$, and in the $S$ region with modes $q_{1,2}^{\pm}$ (solid and dashed red lines). For $k_{y}=0$, the colored regions ($\text{I} \text{ to } \text{V}$) denote distinct transport regimes characterized by different combinations of propagating and evanescent modes as a function of energy. Panel (b) shows a schematic representation of a wave propagating through a {\it perfect} resonant mode (dashed orange) in the {\it phase-matching cavity} of region $\text{I}$. Transmission probabilities in panel (c) for the $T^{-}_{-}$ channel and in panel (d) for the $T^{+}_{+}$ channel are shown for $L=20\text{ nm}$. The color bars between plots correspond to the regions defined in panel (a), and the dashed orange lines in panel (d) indicate the {\it perfect resonances}. Orange insets in panel (d) illustrate the resonant states inside the barrier for incident energies in region~I and different values of the barrier width $L$ and height $V_{0}$. Additional inset in (d) displays an enlarged zoom of a region with nearly zero transmission (blue line). In panels (a), (c) and (d) we use $V_{0}=0.6\text{ eV}$. The insets in panel (d) use the values indicated in each case.}
  \label{fig:2}
\end{figure*}

\section{Phase-Matching Cavity Resonances and Cloaking Effects}

Normal incidence provides a particularly transparent setting to analyze the interplay between propagating and evanescent modes in BG~\cite{Sanderson2013Klein,van2013four,HassaneSaley2022Klein,DellAnna2018From}, multilayer graphene~\cite{Van2013Multiband}, superlattice structures~\cite{Garca2022Generalized} and twisted moiré systems~\cite{He2013Chiral,Handschin2016Fabry}. In BG, the full four-band structure allows internal modes to exist inside an electrostatic barrier that can be selectively decoupled from incident propagating states~\cite{Lee2016Evidence}. This mode-selective decoupling leads to a strong suppression of transmission, consistent with the perfect reflection predicted by effective two-band models, despite the presence of available internal solutions~\cite{Gu2011Chirality,Shytov2015Cloaked,Huang2025Evanescent}. In the present work, we describe this behavior in terms of channel-selective coupling between external propagating states and internal barrier modes, which provides a natural organizing principle for BG transport. Away from strict normal incidence, or outside the corresponding energy window, symmetry constraints are relaxed and additional transport channels or mode mixing can become active~\cite{Gu2011Chirality,Park2011Pi}.

To analyze these effects, we focus on the barrier region $S$. Here, the spectrum follows from Eq.~\ref{Eq: Epristine} with the energy shifted by $V_0$, with corresponding wavenumbers
\begin{equation}
q^{\pm}=\sqrt{(E-V_0)^2\pm\gamma_1|E-V_0|-k_y^2}.
\label{eq: WavenumberQ}
\end{equation}
Equation~\eqref{eq: WavenumberQ} determines which internal solutions are propagating or evanescent and therefore underpins the classification of transport regimes. In the following, we restrict to normal incidence ($k_y=0$), although the mode structure extends straightforwardly to finite transverse momentum~\cite{Van2016Transport}.

Figure~\ref{fig:2}a) shows the superposed band structures of the unperturbed $N$ region (blue) and the barrier $S$ region (red). For a given incident energy $E$, an incoming wave with longitudinal wavenumber $k^{\pm}$ couples at the interfaces to internal solutions with wavenumbers $q_{1,2}^{\pm}$, where the lower index labels the band. Distinct transport regimes arise depending on which of these internal solutions are propagating and, crucially, which are coupled to the incident channel. The five transport regimes and the corresponding propagating and evanescent character of the modes at normal incidence are summarized in Table~\ref{tab:modes}.

As summarized in Table~\ref{tab:modes}, for incidence in region~I (orange in Fig.~\ref{fig:2}a)), electrons outside the barrier connect to hole-like solutions inside the barrier. In the incident region, only the $k^{+}$ mode is propagating, since $k^{-}$ is evanescent (see also Table~\ref{tab:modes}). In this regime, $E<V_0\pm\gamma_1$, and the positive solutions of Eq.~\eqref{eq: WavenumberQ} reduce to $q_1^{\pm}=\sqrt{(V_0-E)(V_0-E\pm\gamma_1)}$, so that both $q_1^{-}$ and $q_1^{+}$ are real. Although both internal modes are propagating, the $q_1^{+}$ mode is decoupled (or cloaked) from the incident $k^{+}$ state~\cite{Gu2011Chirality,Huang2025Evanescent}. We can verified this explicitly by evaluating the transmission (see App.~\ref{App: CSingle}), which yields
\begin{equation}
T_+^+=
\frac{1}{\cos^{2}(q_1^\pm L)+\beta_\pm^{2}\sin^{2}(q_1^\pm L)} ,
\label{eq: Tpp}
\end{equation}
with the mismatch parameter
\begin{equation}
\beta_\pm =
\frac{(q_1^\pm)^{2}E^{2}+(k^+)^{2}(V_0-E)^{2}}
{2q_1^\pm k^+E(V_0-E)} .
\label{eq:beta}
\end{equation}
Here, $q_1^\pm=q_1^{-}$ for $E<V_0$ and $q_1^\pm=q_1^{+}$ for $E>V_0$. Equations~\eqref{eq: Tpp} and~\eqref{eq:beta} show that the $q_1^{+}$ solution does not enter the transmission for incidence from $k^{+}$ in region~I. Transport therefore proceeds through a single internal channel, $k^{+}\rightarrow q_1^{-}\rightarrow k^{+}$, corresponding to $T^{+}_{+}$. 

Figures~\ref{fig:2}(c) and~\ref{fig:2}(d) show the transmission probabilities for the $T^{-}_{-}$ and $T^{+}_{+}$ channels, respectively. As shown in Fig.~\ref{fig:2}d), transmission in region~I exhibits a series of resonances satisfying the phase-matching condition
\begin{equation}
q_1^{-} L = n\pi,
\qquad n \in \mathbb{Z}.
\label{eq: Perfect}
\end{equation}
These resonances yield unit transmission and arise from constructive phase matching of the non-cloaked internal propagating mode across the barrier. At these discrete energies, the accumulated phase allows the wavefunction to match at both interfaces with complete cancellation of reflection, as illustrated schematically in Fig.~\ref{fig:2}b).

Although the resulting mode matching resembles that of a finite quantum well~\cite{Hund1927} or monolayer graphene~\cite{Chen2009Design}, the mechanism is fundamentally different. The relevant internal solutions are not bound states but propagating hole-like modes embedded in the continuum and selectively coupled to the incident channel. We therefore refer to this phase-matched scattering mechanism as a {\it phase-matching cavity}, namely an effective cavity formed by internal phase coherence within a single non-cloaked internal mode of the barrier, in the absence of true bound states and without activating additional transport channels. By varying either the barrier width $L$ or height $V_0$, the number of solutions of Eq.~\eqref{eq: Perfect} can be tuned; the same number of resonances can also be obtained by jointly varying $L$ and $V_0$, as shown in the insets of Fig.~\ref{fig:2}d). This resonance structure reflects the internal four-band mode content of BG and is not captured by reduced two-band models. It is also distinct from Fabry-Pérot resonances arising from interference between multiple barriers~\cite{Ulloa1990Ballistic}.

\begin{table*}[t]
\centering
\label{tab:modes}
\begin{ruledtabular}
\begin{tabular}{c|cc|cc|ll|ll}
Region & $k^+$ & $k^-$ & $q^+$ & $q^-$ & \multicolumn{2}{c|}{$k^+$ incidence} & \multicolumn{2}{c}{$k^-$ incidence} \\
       & (N)   & (N)   & (S)   & (S)   & Cloaking & Transport & Cloaking & Transport \\
\hline
I $(E<V_0-\gamma_1)$   & P & E & P & P & $q^+$ cloaked & Cavity via $q^-$ & \multicolumn{2}{c}{--- (not available)} \\
II $(V_0-\gamma_1<E<\gamma_1)$  & P & E & P & E & $q^+$ cloaked & Leakage via $q^-$  & \multicolumn{2}{c}{--- (not available)} \\
III $(\gamma_1<E<V_0)$ & P & P & P & E & $q^+$ cloaked & Leakage via $q^-$  & $q^-$ cloaked & Cavity via $q^+$ \\
IV $(V_0<E<V_0+\gamma_1)$  & P & P & P & E & --- & Schr\"odinger-like via $q^+$ & $q^+$ cloaked & Leakage via $q^-$  \\
V $(E>V_0+\gamma_1)$   & P & P & P & P & --- & Schr\"odinger-like & --- & Schr\"odinger-like \\
\end{tabular}
\end{ruledtabular}
Mode classification at normal incidence for the five transport regimes identified in Fig.~\ref{fig:2}. Propagating (P) and evanescent (E) character is indicated for each mode and in each region, N and S.  The cloaking and transport columns distinguish incidence from the $k^+$ and $k^-$ channels separately.
\end{table*}

In region~II (yellow in Fig.~\ref{fig:2}), the internal mode $q_1^{-}$ becomes evanescent because $E>V_0-\gamma_1$, while $E<V_0$ still holds. The $q_1^{+}$ mode remains propagating inside the barrier and coexists with the evanescent $q_1^{-}$ solution. However, Eq.~\eqref{eq: Tpp} shows that the propagating $q_1^{+}$ mode remains cloaked for incidence from $k^{+}$~\cite{Gu2011Chirality,Huang2025Evanescent}. As a result, transmission is strongly suppressed in region~II~\cite{Katsnelson2006Chiral}. Within the full four-band description, the evanescent $q_1^{-}$ solution provides a small leakage channel, leading to a finite but very small transmission of order $10^{-4}$, as shown in the inset of Fig.~\ref{fig:2}d). This yields imperfect reflection while preserving cloaking (or decoupling) of the propagating internal mode.

In region~III (light blue in Fig.~\ref{fig:2}), as indicated in Table~\ref{tab:modes}, the $T^{+}_{+}$ channel remains cloaked and continues to exhibit suppressed transmission. The essential difference from region~II is the appearance of an additional propagating incident mode: $k^{-}$ becomes real when $E>\gamma_1$. This activates the transmission channel $T^{-}_{-}$, whose functional form is identical to Eq.~\eqref{eq: Tpp} under the replacement $k^{+}\rightarrow k^{-}$, with $q^\pm=q^{+}$ for $E<V_0$ and $q^\pm=q^{-}$ for $E>V_0$. Note that each $q$ has a band index depending on energy.

Two consequences follow. First, in this energy range (region~III) the internal mode $q_1^{-}$ is evanescent (and cloaked) and does not contribute to the $T^{-}_{-}$ channel. Second, the propagating mode $q_1^{+}$, which is cloaked for incidence from $k^{+}$, becomes the non-cloaked internal channel for incidence from $k^{-}$. As a result, while $T^{+}_{+}$ remains suppressed, $T^{-}_{-}$ supports a series of pronounced phase-matching cavity resonances, as shown in Fig.~\ref{fig:2}c). The transmission process in region~III is therefore $k^{-}\rightarrow q_1^{+}\rightarrow k^{-}$, and the associated perfect resonances satisfy the same phase-matching condition given in Eq.~\eqref{eq: Perfect}. This complementary channel selectivity is a direct consequence of the full four-band structure of BG.

\begin{figure*}
  \centering 
  \includegraphics[width = 0.99 \textwidth]{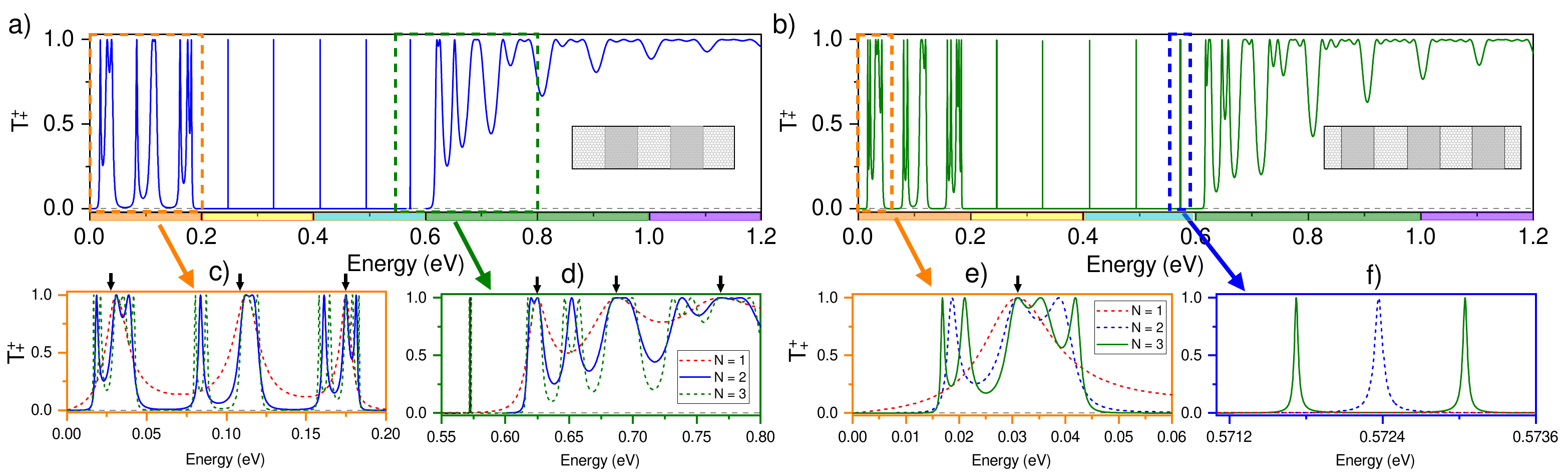}
  \caption{{\bf Multibarrier transport.} Transmission probability in the $T^{+}_{+}$ channel for (a) a double-barrier structure (blue curve) and (b) a triple-barrier structure (green curve). Insets in panels (a) and (b) show schematics of the corresponding multibarrier geometries. Panels (c) and (d) display enlarged views of the transmission spectrum in selected energy windows of panel (a), corresponding to regions~I (orange) and~IV (green), respectively. Panels (e) and (f) show the transmission in additional representative regions of panel (b), highlighting the evolution of the resonance structure with the number of barriers. For comparison, the transmission through a single barrier is shown in red, while blue and green correspond to two and three barriers, respectively. Black arrows in the lower panels indicate the perfect resonances associated with internal phase matching within each barrier. We use the parameters $L = 20\text{ nm}$ for each barrier and separation, and $V_{0} = 0.6 \text{ eV}$.}
  \label{fig:3}
\end{figure*}

For incident energies above the barrier ($E>V_0$), the transmission in the $T^{+}_{+}$ channel approaches the Schr\"odinger-like regime~\cite{Tsu1973Tunneling}, while the corresponding behavior in the $T^{-}_{-}$ channel is recovered for $E>V_0+\gamma_1$. At normal incidence, the scattering channels $T^{+}_{-}$ and $T^{-}_{+}$ vanish due to the absence of mode mixing. For finite transverse momentum $k_y$, however, Eq.~\eqref{eq: WavenumberQ} allows mode mixing and yields nonzero transmission in these channels~\cite{van2013four}. Finally, $T^{-}_{-}$ is identical to $T^{+}_{+}$ up to an energy shift set by the interlayer coupling $\gamma_1$, so that at normal incidence the total conductance receives contributions only from $T^{+}_{+}$ and $T^{-}_{-}$.

\section{Multibarrier Effects: Fabry-Pérot and Perfect Resonances}

We now examine how channel-selective transport and internal mode structure evolve in multibarrier geometries. The central issue is whether the cloaking mechanism identified for a single electrostatic barrier survives when additional interfaces introduce new interference paths. By analyzing double- and triple-barrier configurations within the full four-band framework, we show that the essential physics remains unchanged: at normal incidence, transport is still governed by a single non-cloaked internal channel, while complementary internal modes remain decoupled. Multibarrier structures therefore do not destroy cloaking but instead superimpose additional interference effects on an underlying channel-selective transmission mechanism. In the following, we consider double- and triple-barrier configurations, shown in Fig.~\ref{fig:3}a) and b), respectively. All barriers have identical width $L$ and are separated by intermediate regions of the same length.

For a double-barrier structure consisting of two identical electrostatic barriers, each of width $L$, separated by an intermediate region of equal width $L$, an analytical expression for the transmission probability at normal incidence can be obtained (see App.~\ref{App: DDoble} for further details). When the barrier widths or separation lengths are not equal, the analytical treatment becomes cumbersome, and it is more practical to evaluate the transmission numerically using the general transfer-matrix formalism. For energies below the barrier height, $E<V_0$, the transmission in the $T^{+}_{+}$ channel reads
\begin{equation}
T^+_+ = \frac{256 K^{4} Q^{4}}{A_{0}+\mathcal{R}\cos(2 k^{+} L - \Phi)} ,
\label{eq: TDobleMain}
\end{equation}
where $K = k^{+}(V_0 - E)$ and $Q = q_1^{-} E$. The functions $A_0(k^{+}, q_1^{-})$, $\mathcal{R}(k^{+}, q_1^{-})$ and $\Phi(k^{+}, q_1^{-})$ encode interference effects involving modes in the $N$ and $S$ regions and are given explicitly in App.~\ref{App: DDoble}. For energies above the barrier, the same expression applies upon replacing $q_1^{-}\rightarrow q_1^{+}$ and corresponding band indexes.

Equation~\eqref{eq: TDobleMain} shows that the channel-selective structure identified for a single barrier persists when additional barriers are introduced. At normal incidence, transmission is mediated by a single effective internal channel, while the complementary internal solution remains decoupled from the incident state, independently of whether it is propagating or evanescent. This robustness follows directly from the block-diagonal structure of the transfer matrix at normal incidence, shown in in App.~\ref{App: Smooth}, which preserves channel decoupling as interfaces are added. Consequently, the classification of transport regimes based on channel selectivity extends naturally to multibarrier systems, and cloaking together with imperfect reflection remains robust as the number of barriers increases~\cite{Lamas2024Persistent}.

A direct consequence of channel-selective decoupling is the strong suppression of transmission at normal incidence, a phenomenon often discussed in the literature under the label of anti-Klein tunneling~\cite{VanDuppen2013Klein,Varlet2014Fabry}. Early interpretations associated the appearance of resonant features in multibarrier transmission spectra with a breakdown of cloaking~\cite{Lu2015Destruction}. Subsequent work clarified that resonances do not restore coupling between cloaked internal modes and external states and that cloaking remains operative even when finite transmission appears at resonance~\cite{Lamas2024Persistent}. Within the channel-resolved framework adopted here, this distinction becomes explicit: cloaking corresponds to the decoupling of a specific internal mode from a given incident channel, whereas resonances arise from phase matching within the complementary, non-cloaked internal channel. When this coupled internal mode simultaneously supports evanescent solutions and phase-matched propagation, their combined contribution produces leakage across the barrier and results in imperfect reflection~\cite{Dragoman2008Evidence}. In this sense, the apparent suppression of perfect reflection does not originate from a reactivation of the cloaked mode, but from transport mediated by the single coupled channel through a combination of evanescent tunneling and resonant transmission.

Beyond this qualitative picture, Eq.~\eqref{eq: TDobleMain} reveals an additional robust feature. As in the single-barrier case, the perfect-resonance condition given in Eq.~\eqref{eq: Perfect} is preserved. These perfect resonances persist as the number of barriers increases, and their energy positions remain fixed, as shown in Fig.~\ref{fig:3}. As a result, a multibarrier structure becomes effectively {\it transparent} to an incident wave whose energy satisfies the internal phase-matching condition, in the sense that additional barriers do not modify the transmission at resonance. This transparency is governed by phase coherence within the non-cloaked internal channel rather than by interference between multiple interfaces~\cite{Alvarado2022Biperiodic, Iogansen1987}. Crucially, it requires all barriers to satisfy the same phase-matching condition; a deviation in any single barrier breaks coherence and suppresses perfect transmission.

A second family of resonances originates from interference in the intermediate regions between barriers. As shown in App.~\ref{App: DDoble}, the denominator of Eq.~\eqref{eq: TDobleMain} can be written in terms of a single phase-dependent contribution $\mathcal{R}\cos(2 k^{+} L - \Phi)$, where $2 k^{+} L$ is the phase accumulated across the intermediate region and $\Phi$ encodes reflection phase shifts determined by the internal barrier modes. Transmission maxima therefore occur when
\begin{equation}
2 k^{+} L - \Phi = (2m-1)\pi,
\qquad
m\in\mathbb{Z},
\label{eq: General2ResoMain}
\end{equation}
corresponding to Fabry-Pérot-like constructive interference.

\begin{figure*}
  \centering 
  \includegraphics[width = 0.95 \textwidth]{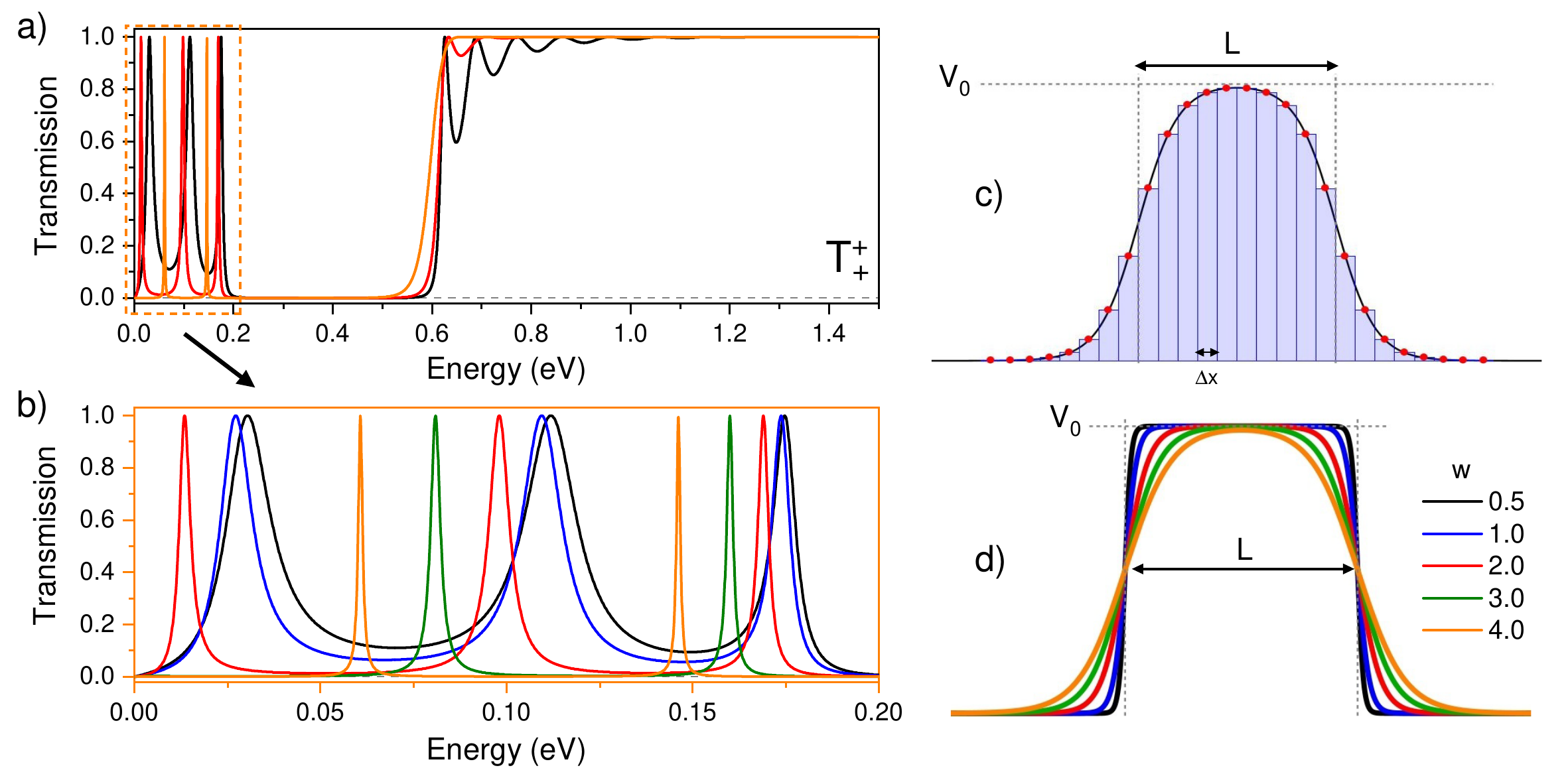}
 \caption{{\bf Smooth barrier profiles.} 
(a) Transmission probability in the $T^{+}_{+}$ channel for a single barrier with a smooth profile as a function of energy, shown for different values of the characteristic smoothing length $w$. 
(b) Enlarged view of panel (a) in Region I, including additional transmission curves for $w=\{0.5,1.0,2.0,3.0,4.0\}$, with the corresponding color scale indicated in panel (d). 
(c) Schematic discretization of the smooth barrier used in the transfer-matrix calculation. 
(d) Barrier potential profile for representative values of the smoothing length $w$. Color scheme of (a), (b) is indicated in (d).}
\label{fig:4}
\end{figure*}

Figure~\ref{fig:3} illustrates this hierarchy for systems with one, two, and three barriers. A single barrier supports isolated perfect resonances determined by Eq.~\ref{eq: Perfect}. Introducing a second barrier preserves the position of each perfect resonance and generates two additional resonances that flank it symmetrically, forming a characteristic triplet structure. Adding a third barrier further splits these side resonances, resulting in five resonances in total. More generally, for a system of $N$ identical barriers, each perfect resonance is accompanied by $N-1$ additional resonances on each side, giving a total of $2N-1$ resonances. This hierarchy reflects the increasing number of interfering paths in the intermediate regions, while the perfect resonance itself remains invariant. A similar splitting mechanism has been reported for monolayer graphene~\cite{Xu2014Resonant,Xu2019Resonant}.

This structure highlights the distinct physical origins of the two contributions to the transmission spectrum. Perfect resonances are governed solely by internal phase matching within individual barriers, whereas the surrounding resonances arise from Fabry-Pérot-like interference between adjacent barriers. Increasing the number of barriers therefore enriches the resonance structure without altering the energy or the nature of the perfect resonances. Unlike the perfect resonances fixed by Eq.~\ref{eq: Perfect}, the Fabry-Pérot-like resonances do not enforce exact cancellation of reflections at the interfaces, so that transmission is enhanced but does not generally reach unity. These interference-induced resonances coexist with cloaking of internal modes without restoring coupling to the cloaked channel. Additional transmission peaks also appear in other energy regions due to inter-barrier interference, including regions~II and~III where a single barrier would otherwise exhibit strongly suppressed transmission. These peaks correspond to conventional interference-induced resonances, analogous to those observed in semiconductor heterostructures~\cite{Garcia1993Descripcion,Garcia1994Descripcion}.

\begin{figure*}
  \centering 
  \includegraphics[width = 0.85 \textwidth]{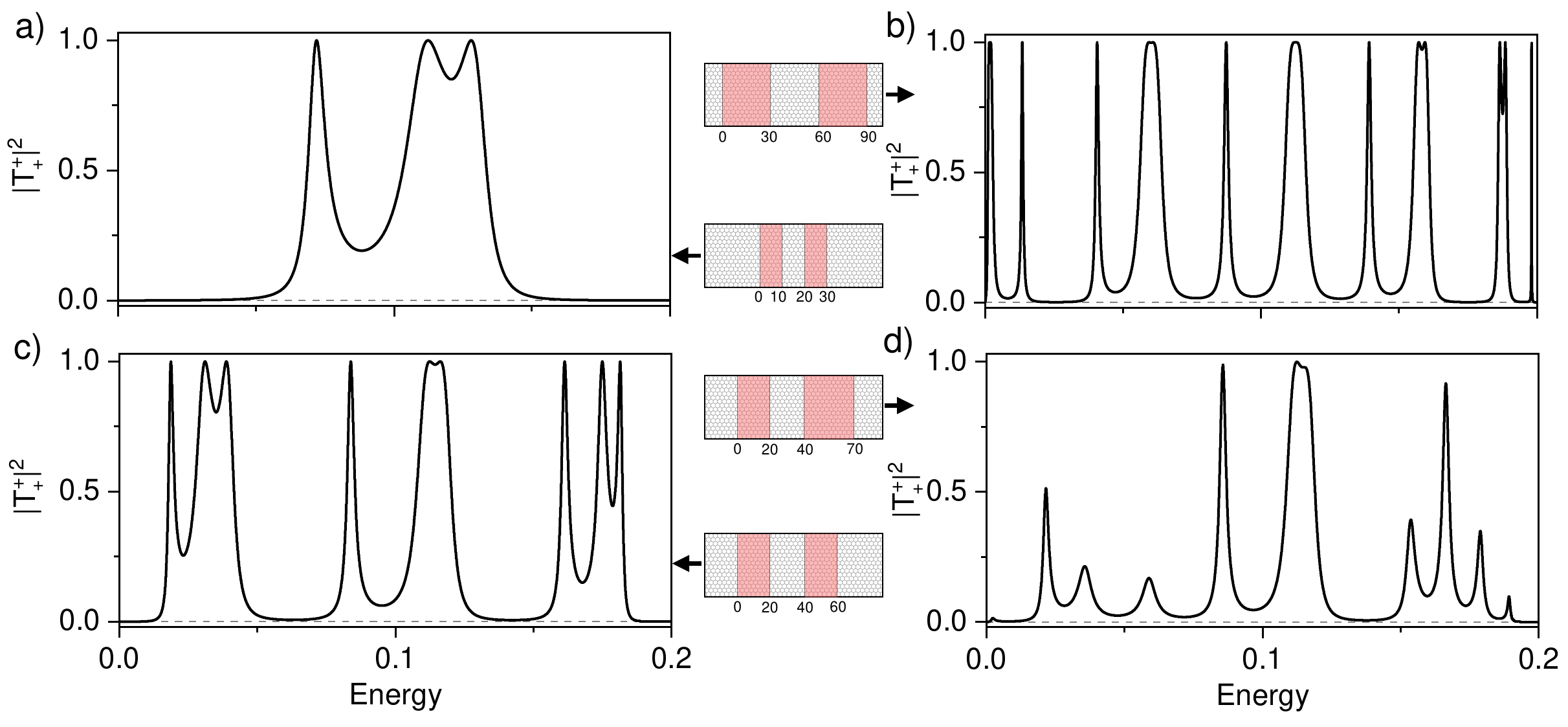}
  \caption{{\bf Geometry Effects.} Transmission probability in the $T^{+}_{+}$ channel for a double-barrier structure with different barrier and separation lengths (in nm). The corresponding geometry and dimensions are shown schematically in each panel. The energy range lies within region I. In all panels, we use $V_0 = 0.6,\mathrm{eV}$.}
  \label{fig:5}
\end{figure*}

\section{Robustness of the Phase-Matching Cavity to Smooth Barriers}
In realistic experimental devices, electrostatic barriers are not perfectly abrupt but exhibit smooth edges due to the finite screening length of the gate potential and the spatial extent of the gate geometry~\cite{Baringhaus2015Ballistic,Huard2007Transport}. To assess the robustness of the phase-matching cavity mechanism discussed in the previous sections, we model the barrier profile as
\begin{equation}
V(x) = \frac{V_0}{2}\left(\tanh\frac{x}{w} - \tanh\frac{x - L}{w}\right),
\label{eq:VtanhMain}
\end{equation}
where $L$ is the width and $w$ controls the smoothing of the edges. As $w$ increases, the effective plateau region is slightly reduced, modifying the accumulated internal phase. The potential increases from approximately $10\%$ to $90\%$ of its maximum value over a distance of about $2.2w$, and the abrupt-barrier limit analyzed above is recovered for $w \to 0$. The smooth profile is discretized into $N$ consecutive slabs of equal width and treated within the same transfer-matrix framework described previously (see also App.~\ref{App: Smooth}). 

Figure~\ref{fig:4}(a) shows the transmission probability in the $T^{+}_{+}$ channel for different values of the smoothing parameter $w$, while Fig.~\ref{fig:4}(b) focuses on region I, where the perfect resonances were identified for abrupt barriers. Fig.~\ref{fig:4}(c) illustrate the discretization of the potential profile shown in Fig.~\ref{fig:4}(d). As $w$ increases, the resonance peaks shift systematically toward lower energies. For moderate smoothing, the first few resonances are displaced but remain clearly visible. For sufficiently large $w$, some resonances are progressively suppressed. This evolution reflects the modification of the accumulated internal phase caused by the smoother barrier edges. Since the perfect transmission condition is governed by phase matching inside the barrier, changes in the spatial profile alter the effective phase accumulation and therefore shift the resonance energies. 

As mentioned before, this behavior closely resembles that of a conventional quantum well. In a square well, the bound-state energies depend sensitively on the shape and depth of the confining potential; if the profile is gradually deformed, the discrete spectrum shifts and certain states may disappear. In the present case, although the resonances correspond to phase-matching conditions of propagating modes rather than true bound states, their evolution under profile smoothing follows the same geometric principle. Importantly, the resonance structure persists over a broad range of smoothing lengths, demonstrating that the phase-matching cavity mechanism is not an artifact of the abrupt-barrier approximation but a robust consequence of phase accumulation within the barrier region.

A further consequence of the smooth-barrier geometry concerns the structure of the transport channels at normal incidence. As shown in App.~\ref{App: Smooth}, the two channels associated with the $k^+$ and $k^-$ modes remain completely independent of each other, even when the barrier profile varies smoothly in space. This separation follows from an underlying layer symmetry of the Hamiltonian at normal incidence, which ensures that the two channels do not mix anywhere inside the barrier, regardless of how the potential changes from slab to slab. As a result, an electron incident in one channel remains in that same channel throughout the scattering process. The transport problem therefore reduces to two independent single-channel problems. Importantly, this exact decoupling does not rely on the abrupt-barrier approximation. It remains valid for any smooth scalar potential profile $V(x)$, so the single-channel phase-matching condition $qL = n\pi$ identified for sharp barriers continues to govern the perfect resonances in the smooth case.

\section{Dependence on Barrier Width and Separation}
\label{sec:GeometryDependence}
Figure~\ref{fig:5} shows how the barrier widths and inter-barrier separation modify the transmission spectrum in region~I. The geometry has a strong impact on the resonance pattern, including the number, width, and overall line shape of the resonances. As a reference, we consider the symmetric double-barrier case with $L=20\,\mathrm{nm}$ shown in Fig.~\ref{fig:3}. When the barrier width is reduced (Fig.~\ref{fig:5}a), the number of resonances decreases. When the barrier width is increased (Fig.~\ref{fig:5}b), the number of resonances increases, and the resonance line shapes are also modified. For asymmetric barriers (Fig.~\ref{fig:5}d), some resonances disappear because the phase-matching condition is no longer satisfied simultaneously in both barriers. Nevertheless, several resonances persist, consistent with solutions for which the accumulated phase remains compatible with the two different barrier lengths. As discussed before, changing the barrier lengths does not mix the decoupled modes at $k_y=0$, so mode decoupling remains intact. What changes is the accumulated phase, which determines the number and positions of the resonances. This trend supports the phase-matching cavity interpretation: reducing the barrier width reduces the number of resonances, while increasing it produces additional resonances, in direct analogy with the shift and rearrangement of discrete levels when the profile of a conventional quantum well is deformed.

\section{Discussions and Conclusions}

The strong sensitivity of perfect resonances to barrier width, separation, and electrostatic potential highlights both their fragility and their diagnostic value. Within the full four-band, mode-resolved framework developed here, perfect transmission is tied to the phase-matching cavity mechanism: a single non-cloaked internal mode satisfies a phase-matching condition and yields unit transmission governed by internal phase coherence within each barrier, rather than by multireflection effects alone. Channel-selective decoupling, which underlies cloaking, corresponds to the symmetry-protected isolation of a specific internal barrier mode from a given external incident channel and remains operative regardless of whether the internal solution is propagating or evanescent. As a result, resonant peaks enhance transport through allowed channels without restoring access to cloaked modes, while residual leakage reflects imperfect reflection mediated by evanescent contributions rather than a breakdown of cloaking.

These conclusions are made explicit by the analytical solutions obtained for single- and double-barrier geometries, where the transmission naturally separates into two distinct families of resonances. Perfect resonances originate from internal phase matching within individual barriers and obey the condition in Eq.~\eqref{eq: Perfect}, whereas Fabry-Pérot-like resonances arise from interference between barriers and are governed by Eq.~\eqref{eq: General2ResoMain}. Although ballistic transport through electrostatic barriers in bilayer graphene has been studied extensively~\cite{van2013four,Xu2014Resonant,Xu2019Resonant}, previous work did not clearly disentangle mode-selective decoupling from resonance-induced finite transmission in multibarrier systems. The present analysis provides this separation within a unified four-band description.

Beyond analytically tractable single- and double-barrier cases, we employed a general matrix mode-matching approach to analyze multibarrier structures with an arbitrary number of barriers. While the growing number of scattering paths precludes a compact analytical resonance condition for large $N$, numerical evaluation of the full transfer matrix shows that the channel-selective structure identified throughout the paper is preserved. In particular, perfect resonances associated with internal phase matching remain fixed in energy as the number of barriers increases, while additional interference-induced resonances proliferate around them. This confirms that the separation between perfect and Fabry-Pérot-like resonances remains valid beyond the simplest geometries.

The mechanisms identified here are not specific to bilayer graphene. They rely on the coexistence of multiple internal modes and on multiband or effectively parabolic dispersions near the transport energy. Channel-selective decoupling, phase-matching cavity formation, and the coexistence of perfect and interference-induced resonances are therefore expected to arise in other multiband two-dimensional systems, provided that electrostatic barriers couple selectively to internal modes~\cite{Campos2012Quantum,Pereira2015Landau}.

Our results also provide a microscopic perspective on recent Corbino-geometry experiments reporting conductance signatures consistent with angularly selective tunneling and tunneling suppression in BG~\cite{Elahi2024Direct}. In those measurements, the observable is the total conductance, integrated over all incident angles and transport channels, so individual mode contributions cannot be directly resolved. However, the ray-tracing arguments in Ref.~\cite{Elahi2024Direct} indicate that a subset of trajectories crosses the barrier region, implying that the barrier is not uniformly opaque. Within the framework developed here, such trajectories do not require a reactivation of cloaked channels and can be understood in terms of resonance-assisted transmission through non-cloaked internal modes. In this sense, the phase-matching cavity mechanism and the Fabry-Pérot-like resonances provides a natural microscopic interpretation of how strong conductance modulations and tunneling suppression can coexist with internal propagating solutions inside electrostatic barriers. The persistence of perfect resonances under smooth deformations of the barrier profile further confirms that the phase-matching cavity is governed by accumulated internal phase rather than by the presence of sharp interfaces. Even when the electrostatic potential varies continuously in space, transport at normal incidence remains effectively single-channel and the resonance condition continues to be set by the phase acquired across the barrier. The phase-matching cavity is therefore a robust interference mechanism intrinsic to the multiband structure of bilayer graphene.

Finally, these results have direct experimental implications. For a single electrostatic barrier, small variations in barrier width or potential mainly shift the energies at which perfect resonances occur. In contrast, in structures with multiple barriers, such variations break the phase matching required across different barriers and can strongly reduce or eliminate perfect resonances. As a result, perfectly resonant transport becomes difficult to observe in devices with non-identical barriers. By contrast, the observation of stable and repeatable perfect resonances may provide a clear indication of barrier uniformity and offer a sensitive way to probe channel-selective transport in multibarrier graphene devices.

\section*{ACKNOWLEDGMENTS}
We thank Francisco Guinea, Federico Escudero and Ramon Carrillo-Bastos for useful discussions. D.L acknowledges the support from the China Scholarship Council (CSC) program, project ID: 202406960092, and by the Natural Science Foundation of Shaanxi Province, Grant No.\ 2025JC-YBQN-023. P.A.P acknowledged support from the “Severo Ochoa” Programme for Centres of Excellence in R\&D (CEX2020-001039-S/AEI/10.13039/501100011033) financed by MICIU/AEI/10.13039/501100011033 and from NOVMOMAT, Grant PID2022-142162NB-I00 funded by MCIN/AEI/ 10.13039/501100011033 and, by "ERDF A way of making Europe". P.A.P acknowledges funding by Grant No.\ JSF-24-05-0002 of the Julian Schwinger Foundation for Physics Research.

\appendix

\section{Transmission Matrix Method}
\label{App: AMatrix}
We consider AB-stacked BG subject to electrostatic barriers along the propagation direction $x$. The system is assumed to be translationally invariant along the transverse $y$ direction. Following standard approaches developed for semiconductor heterostructures~\cite{Yamamoto1989Transmission,Wu1991Quantum,Ulloa1990Ballistic}, we introduce $n$ electrostatic barriers aligned along $x$. Such barriers have been experimentally realized in both monolayer graphene~\cite{Young2009Quantum,Stander2009Evidence} and BG~\cite{Mylnikov2022Terahertz,Titova2023Ultralow}. The simplest configuration corresponds to a piecewise-constant on-site electrostatic potential $V_0$, as discussed in the main text. For a single barrier, the system is divided into three regions: an unperturbed left region ($x<0$), an electrostatically modified central region ($0<x<L$), and an unperturbed right region ($x>L$). The unperturbed regions are denoted as $n$, while the central region is denoted as $S$. More generally, we consider symmetric multibarrier structures in which each barrier and intermediate region has the same width $L$, forming a periodic superlattice along $x$~\cite{Wu1991Quantum}. 

For convenience, we introduce a characteristic length $l_0=100$ nm and the corresponding energy scale $E_0=\hbar v_f/l_0 \approx 6.58$ meV. All energies and wavenumbers are expressed in dimensionless form via
\begin{equation}
V_0 \rightarrow V_0/E_0, \quad
E \rightarrow E/E_0, \quad
\gamma_1 \rightarrow \gamma_1/E_0, \quad
k \rightarrow l_0 k.
\end{equation}
Because the system is invariant along $y$, the transverse momentum $k_y$ is conserved. The wavefunction can therefore be written as $\Psi(x,y)=\Phi(x)e^{ik_y y}$, where
\begin{equation}
\Phi(x)=
\left( \phi_{A_1}(x), \phi_{B_1}(x), \phi_{B_2}(x), \phi_{A_2}(x) \right)^{T}.
\end{equation}
We focus here on normal incidence, $k_y=0$, for which the problem reduces to one dimension along $x$. Extensions to finite incidence angles follow straightforwardly~\cite{van2013four,Van2016Transport}. Dropping the explicit $x$ dependence for compactness, the Schr\"odinger equation yields the coupled first-order equations
\begin{align}
-i \frac{d\phi_{B_1}}{dx} &= (E-V_0)\phi_{A_1} - \gamma_1 \phi_{B_2}, \\
-i \frac{d\phi_{A_1}}{dx} &= (E-V_0)\phi_{B_1}, \\
-i \frac{d\phi_{A_2}}{dx} &= (E-V_0)\phi_{B_2} - \gamma_1 \phi_{A_1}, \\
-i \frac{d\phi_{B_2}}{dx} &= (E-V_0)\phi_{A_2}.
\end{align}
Eliminating auxiliary components, one finds that $\phi_{A_1}$ satisfies
\begin{equation}
\frac{d^2 \phi_{A_1}}{dx^2} = (q^{\pm})^2 \phi_{A_1},
\end{equation}
with
\begin{equation}
(q^{\pm})^2 = (E-V_0)^2 \pm \gamma_1 |E-V_0|.
\end{equation}
An analogous structure holds for the remaining components. In the $N$ regions ($V_0=0$), the corresponding wavenumbers reduce to
\begin{equation}
(k^{\pm})^2 = E^2 \pm |E| \gamma_1.
\end{equation}
The general solution in a region with constant $V_0$ is therefore a superposition of four plane waves. For example,
\begin{equation}
\phi_{A_1}(x) =
A e^{i q^{+} x} + B e^{-i q^{+} x}+C e^{i q^{-} x} + D e^{-i q^{-} x},
\end{equation}
with analogous expressions for the other components. Collecting terms, the wavefunction in $S$ region can be written as
\begin{equation}
\Phi_{II}(x) = \Omega_{II} P_{II}(x)
\begin{pmatrix}
A \\ B \\ C \\ D
\end{pmatrix},
\end{equation}
where 
\begin{equation}
\Omega_{II}=\begin{pmatrix}
  1& 1 &1  &1 \\
  d^{+}_{+}&d^{+}_{-}  &d^{-}_{+}  &d^{-}_{-} \\
 h^{+}& h^{+} &h^{-}  &h^{-} \\
 d^{+}_{+}h^{+} &d^{+}_{-}h^{+}  &d^{-}_{+}h^{-} &d^{-}_{-}h^{-}
\end{pmatrix},
\end{equation}
with $d^s_\pm=\pm \frac{q^s}{E-V_0}$, with $s=\pm1$, $h^\pm=\mp\text{sign}(E-V_0)$ and $P_{II}(x)$ contains the phase factors. Explicit expressions are given by
\begin{equation}
P_{II}(x)=\mathrm{diag}
\left(e^{i q^{+} x},e^{-i q^{+} x},e^{i q^{-} x},e^{-i q^{-} x}\right),
\end{equation}
with coefficients defined consistently with the main text. In the $N$ regions, the eigenstates are obtained by setting $V_0=0$, yielding analogous matrices $\Omega_{\mathrm{I(III)}}$ and $P_{\mathrm{I(III)}}(x)$ with $q^{\pm}$ replaced by $k^{\pm}$. The wavefunction in the left lead ($N$) contains incoming and reflected components,
\begin{equation}
\Phi_{\mathrm{I}}(x)=\Omega_{\mathrm{I}} P_{\mathrm{I}}(x)
\begin{pmatrix}
\delta_{s,1} \\
r^{s}_{+} \\
\delta_{s,-1} \\
r^{s}_{-}
\end{pmatrix},
\end{equation}
while in the right lead only transmitted waves are present,
\begin{equation}
\Phi_{\mathrm{III}}(x)=\Omega_{\mathrm{III}} P_{\mathrm{III}}(x)
\begin{pmatrix}
t^{s}_{+} \\
0 \\
t^{s}_{-} \\
0
\end{pmatrix}.
\end{equation}
Continuity of the wavefunction at each interface relates these coefficients via transfer matrices. Defining the interface matrices
\begin{equation}
\begin{aligned}
M_{\mathrm{I}\to\mathrm{II}} &= P_{\mathrm{I}}^{-1}(0)\,\Omega_{\mathrm{I}}^{-1}\,\Omega_{\mathrm{II}}\,P_{\mathrm{II}}(0), \\
M_{\mathrm{II}\to\mathrm{III}} &= P_{\mathrm{II}}^{-1}(L)\,\Omega_{\mathrm{II}}^{-1}\,\Omega_{\mathrm{III}}\,P_{\mathrm{III}}(L).
\end{aligned}
\label{eq: Connection}
\end{equation}
the total scattering matrix is
\begin{equation}
S = M_{\mathrm{I}\to\mathrm{II}} M_{\mathrm{II}\to\mathrm{III}}.
\end{equation}
By writing $\Delta = S_{11}S_{33} - S_{13}S_{31}$, the transmission amplitudes are
\begin{equation}
\begin{aligned}
t^{+}_{+} &= \frac{S_{33}}{\Delta}, &
t^{+}_{-} &= -\frac{S_{31}}{\Delta}, \\
t^{-}_{+} &= -\frac{S_{13}}{\Delta}, &
t^{-}_{-} &= \frac{S_{11}}{\Delta}.
\end{aligned}
\label{eq:tamplitudes}
\end{equation}
with reflection amplitudes obtained analogously. Transmission and reflection probabilities are computed from the longitudinal current density, $\mathbf{J} = v_f \Psi^\dagger \boldsymbol{\sigma} \Psi$. This allows one to define channel-resolved transmission and reflection coefficients as
\begin{equation}
\begin{alignedat}{2}
T^{s}_{s'} &= \frac{\left|\mathbf{J}^{\,s'}_{\mathrm{tra}}\right|}
{\left|\mathbf{J}^{\,s}_{\mathrm{in}}\right|},
\qquad
&R^{s}_{s'} &= \frac{\left|\mathbf{J}^{\,s'}_{\mathrm{ref}}\right|}
{\left|\mathbf{J}^{\,s}_{\mathrm{in}}\right|}.
\end{alignedat}
\label{eq:TRcurrent}
\end{equation}
For propagating modes, these expressions reduce to
\begin{equation}
\begin{aligned}
T^{s}_{s'} &= \frac{k^{s'}}{k^s}\, |t^{s}_{s'}|^2, \\
R^{s}_{s'} &= \frac{k^{s'}}{k^s}\, |r^{s}_{s'}|^2 .
\end{aligned}
\label{eq:TRcoeff}
\end{equation}
where $k^s$ and $k^{s'}$ denote the longitudinal wavevectors of the incident and outgoing channels, respectively, and $t^{s}_{s'}$ and $r^{s}_{s'}$ are the corresponding transmission and reflection amplitudes. The prefactor $k^{s'}/k^s$ accounts for the ratio of group velocities between outgoing and incoming modes and ensures proper current normalization. Probability conservation then requires 
\begin{equation}
\sum_{s'} \left( T^{s}_{s'} + R^{s}_{s'} \right) = 1    
\end{equation}
for each incident channel $s$.

\section{Extension to multibarrier systems}
\label{App: BMulti}
The extension to multibarrier systems follows the standard transfer-matrix construction used for one-dimensional superlattices~\cite{Yamamoto1989Transmission,Wu1991Quantum,Ulloa1990Ballistic}. The structure is divided into regions of constant electrostatic potential, and the wavefunction is matched continuously at each interface.
We consider $n$ identical electrostatic barriers of width $L$, separated by $n-1$ intermediate regions of the same width $L$. The multibarrier region therefore consists of $2n-1$ segments of equal length $L$, alternating between barrier and intermediate regions. The interfaces are located at positions $x_j = (j-1)L$, with $j=1,\dots,2n-1$.
In each segment the wavefunction is written as a superposition of four plane-wave solutions, whose amplitudes are collected into the coefficient vector $\mathbf{C}_j = (A_j,B_j,C_j,D_j)^T$.
At the left boundary $x=0$, the incoming and reflected amplitudes in the left lead are related to the coefficients in the first segment according to
\begin{equation}
\begin{pmatrix}
\delta_{s,1} \\
r^{s}{+} \\
\delta{s,-1} \\
r^{s}_{-}
\end{pmatrix}=
M_0
\begin{pmatrix}
A_1 \\
B_1 \\
C_1 \\
D_1
\end{pmatrix},
\end{equation}
where
\begin{equation}
M_0 = P_{1}^{-1}(0)\Omega_I^{-1}\Omega_{II}P_{II}(0).
\end{equation}
Continuity of the wavefunction at each internal interface $x=x_j$ relates the coefficients in adjacent segments according to
\begin{equation}
\mathbf{C}_j =P_j^{-1}(x_j)\Omega_j^{-1}\Omega_{j+1}P_{j+1}(x_j)\mathbf{C}_{j+1},
\end{equation}
 where  j=$1,\dots,2n-2$. The transfer matrix across the internal multibarrier region is obtained by multiplying the interface matrices in sequence,
\begin{equation}
M_T =
\prod_{j=1}^{2n-2}\left[P_j^{-1}(x_j)\Omega_j^{-1}\Omega_{j+1}P_{j+1}(x_j)\right],
\end{equation}
where the product is ordered along the propagation direction.
At the right boundary $x=(2n-1)L$, the coefficients in the last segment are matched to the transmitted modes in the right lead through
\begin{equation}
M_{2n-1} =
P_{2n-1}^{-1}[(2n-1)L]\Omega_{2n-1}^{-1}\Omega_{\mathrm{III}}P_{\mathrm{III}}[(2n-1)L].
\end{equation}
The total transfer matrix of the structure is therefore given by
\begin{equation}
S = M_0 M_T M_{2n-1},
\end{equation}
and the transmission and reflection coefficients follow as in the single-barrier case.

\section{Transmission in a Single Barrier}
\label{App: CSingle} 

For a single electrostatic barrier, we follow the same transfer-matrix procedure introduced previously. As discussed in the main text, in the regime relevant to region~I only the external mode $k^{+}$ is propagating, while the solution $k^{-}$ is evanescent and does not contribute to transport. In this case, transmission proceeds through the channel $k^{+}\rightarrow q \rightarrow k^{+}$ and can be written in closed analytical form as
\begin{equation}
T^+_+=
\frac{1}{\cos^{2}(qL)+\beta^{2}\sin^{2}(qL)} .
\label{eq:TSingle}
\end{equation}
The external longitudinal wavenumber is
\begin{equation}
k^{+}=\sqrt{E(E+\gamma_{1})},
\end{equation}
while the internal wavenumbers inside the barrier are
\begin{equation}
q^\pm=\sqrt{(E-V_{0})^{2}\pm\gamma_{1}|E-V_{0}|}.
\end{equation}
Depending on the energy relative to the barrier height, the relevant internal wavenumber entering Eq.~\eqref{eq:TSingle} is defined as
\begin{equation}
q=
\begin{cases}
q^-, & E<V_{0}, \\
q^+, & E>V_{0}.
\end{cases}
\end{equation}
The mismatch parameter $\beta$ appearing in Eq.~\eqref{eq:TSingle} is given by
\begin{equation}
\beta=
\frac{q^{2}E^{2}+(k^{+})^{2}(V_{0}-E)^{2}}
{2qk^{+}E(V_{0}-E)} .
\label{eq:betaSingle}
\end{equation}
Equation~\eqref{eq:TSingle} shows that transmission through a single barrier is governed by the phase accumulated by the internal mode across the barrier region, $qL$. The denominator consists of an oscillatory contribution proportional to $\sin(qL)$ and a background contribution proportional to $\cos(qL)$, reflecting interference between forward- and backward-propagating amplitudes inside the barrier.

\subsection*{Resonances: Single Barrier}
The resonance structure follows directly from Eq.~\eqref{eq:TSingle}. Perfect transmission occurs when the sine term vanishes, this is $
\sin(qL)=0$, i.e, 
\begin{equation}
qL=n\pi,
\qquad
n\in\mathbb{Z}.
\label{eq: perfectreso}
\end{equation}
When this condition is satisfied and $q\neq 0$, Eq.~\eqref{eq:TSingle} yields $T^+_+=1$. These resonances originate from phase matching of the internal mode across the barrier, such that reflections at the interfaces cancel exactly. The barrier therefore behaves as a phase-coherent scattering region supporting perfect transmission. Away from the exact phase-matching condition, Eq.~\eqref{eq:TSingle} also describes the usual Fabry-Pérot-like oscillations arising from partial interference between forward- and backward-propagating waves inside the barrier. 

\section{Transmission in a Double Barrier}
\label{App: DDoble}
For a two-barrier system, the same transfer-matrix procedure applies. The resulting transmission probability for the $T^+_+$ channel at normal incidence can be written as
\begin{equation}
T^+_+ = \frac{256K^{4} Q^{4}}{R^{2} + I^{2}} .
\label{eq:TDoble}
\end{equation}
For convenience, we define $K = k^{+}(V_{0} - E)$ and $Q = q_{1}^{-} E$ for $E<V_{0}$. For energies above the barrier, the replacement $q_{1}^{-}\rightarrow q_{1}^{+}$ is understood throughout. The two terms in the denominator of the above equation are given by
\begin{eqnarray} 
R &=& 16 K^{2} Q^{2} \cos(2 q_{1}^{-} L) \nonumber \\ && -\, 8 (K^{2} - Q^{2})^{2} \sin^{2}(q_{1}^{-} L)\sin^{2}(k^{+} L), \\
I &=& 8 K Q (K^{2} + Q^{2}) \sin(2 q_{1}^{-} L) \nonumber \\ && +\, 4 (K^{2} - Q^{2})^{2} \sin^{2}(q_{1}^{-} L)\sin(2 k^{+} L). \nonumber
\end{eqnarray} This structure makes explicit that transport is governed by interference between phases accumulated in the barrier regions, through $q_{1}^{-}L$, and in the intermediate region, through $k^{+}L$.

\subsection*{Resonances: Double Barrier}
As in the single-barrier case, a first family of resonances corresponds to exact unit transmission and occurs when the internal phase-matching condition in Eq.~\eqref{eq: perfectreso} is satisfied. These resonances are inherited directly from the single-barrier problem and correspond to perfect transmission through each barrier independently. A second family of resonances originates from interference in the intermediate region between the two barriers. Introducing
\begin{eqnarray} 
A &=& 16 K^{2} Q^{2} \cos(2 q_{1}^{-} L) - B,\nonumber \\ 
B &=& 4 (K^{2} - Q^{2})^{2} \sin^{2}(q_{1}^{-} L), \\ 
C &=& 8 K Q (K^{2} + Q^{2}) \sin(2 q_{1}^{-} L), \nonumber 
\end{eqnarray} 
the denominator of Eq.~\ref{eq:TDoble} can be written as 
\begin{eqnarray} R^{2} + I^{2} &=& A_{0} + A_{c}\cos(2 k^{+} L) + A_{s}\sin(2 k^{+} L), 
\end{eqnarray} 
where $A_{0} = A^{2} + B^{2} + C^{2}$, $A_{c} = 2AB$, and $A_{s} = 2CB$. 
Defining $\mathcal{R}=\sqrt{A_{c}^{2}+A_{s}^{2}}$ and $\Phi=\arctan(A_{c},A_{s})$, we get 
\begin{equation} A_{c}\cos(2 k^{+} L) + A_{s}\sin(2 k^{+} L) = \mathcal{R}\cos(2 k^{+} L - \Phi). 
\end{equation} 
Transmission peaks occur when the denominator is minimized, leading to the condition
\begin{equation}
2 k^{+} L - \Phi = (2m-1)\pi,
\qquad
m\in\mathbb{Z}.
\label{eq: General2Reso}
\end{equation}
This condition describes Fabry-Pérot-like resonances associated with constructive interference between multiple reflections in the region separating the two barriers. Unlike the perfect resonances fixed by Eq.~\eqref{eq: perfectreso}, these resonances do not generally yield unit transmission.
\\
\section{Transmission in a Triple Barrier} \label{App: ETriple}
For a three-barrier structure, transport can again be treated within the same transfer-matrix framework. The system consists of three identical barriers of width $L$, separated by two identical intermediate regions of width $L$. The transmission probability can be expressed as
\begin{equation}
T^+_+ = \frac{4096 K^{6} Q^{6}}{R_{3}^{2} + I_{3}^{2}} ,
\label{eq:TTriple}
\end{equation}
where $K$ and $Q$ are defined as in the double-barrier case. The denominator $R_{3}^{2}+I_{3}^{2}$ is given by
\begin{equation}
\Delta=S_{11}S_{33}-S_{13}S_{31},
\end{equation}
with $S$ the total transfer matrix of the three-barrier system. In contrast to the double-barrier geometry, the triple-barrier structure involves two intermediate regions and therefore multiple independent interference phases. As a result, $R_{3}$ and $I_{3}$ are trigonometric polynomials of the phases accumulated both inside the barriers, $q_{1}^{\pm}L$, and in the intermediate regions, $k^{\pm}L$. These contributions arise from multiple scattering paths and cannot be easily reduced to a single Fabry-Pérot phase condition.

\subsection*{Resonances: Triple Barrier}

As in the single- and double-barrier cases, a first family of resonances corresponds to exact unit transmission when the internal phase-matching condition in Eq.~\eqref{eq: perfectreso} is satisfied. These perfect resonances originate from phase coherence within each barrier and persist independently of the number of barriers. In addition, the triple-barrier geometry supports a richer set of resonances associated with constructive interference in the two intermediate regions. These resonances depend on the accumulated phase $k^{+}L$ between successive barriers and are shifted by energy-dependent phase factors arising from multiple reflections at the interfaces. As a consequence, not all resonance conditions lead to unit transmission, reflecting the increased complexity of interference processes in multibarrier structures. Unlike the single- and double-barrier cases, a compact analytical resonance condition is difficult to obtain for the triple-barrier geometry. In practice, the resonance structure is most conveniently analyzed using the general matrix method described in App.~\ref{App: BMulti}.

\section{Extension to smooth potential profiles}
\label{App: Smooth} 
The transfer-matrix construction can be extended to treat arbitrary smooth electrostatic profiles $V(x)$ by means of a staircase discretisation. Consider a potential of the form
\begin{equation}
V(x) = \frac{V_0}{2}\left(\tanh\frac{x}{w} - \tanh\frac{x - L}{w}\right),
\label{eq:Vtanh}
\end{equation}
where $V_0$ is the barrier height, $w$ controls the smoothness of the edges, and $L$ is the nominal barrier length. The barrier region $0 \leq x \leq L$ is partitioned into $N$ consecutive slabs of equal width $\Delta x = L/N$. Slab~$j$ ($j=1,\dots,N$) is centred at $x_j = (j - \tfrac{1}{2})\Delta x$ and assigned the constant potential $V_j = V(x_j)$. The effective energy in slab~$j$ is $\bar{E}_j = E - V_j$, and the longitudinal wavenumbers follow from the same dispersion as in the uniform case,
\begin{equation}
(Q_j^{\pm})^2 = \bar{E}_j^2 \pm \gamma_1|\bar{E}_j|.
\label{eq:Qjsmooth}
\end{equation}
These expressions reduce to the uniform-barrier wavenumbers $q^\pm$ defined in the main text when $V_j$ is constant. The eigenvector matrix $\Omega_j$ and the propagation matrix $P_j(\Delta x)$ in each slab retain the structure given in App.~\ref{App: AMatrix}, with the replacements $E-V_0 \to \bar{E}_j$ and $q^{\pm} \to Q_j^{\pm}$. At the left boundary $x=0$, the interface matrix reduces to $S_0 = \Omega_{\mathrm{I}}^{-1}\,\Omega_1$, since $P_{\mathrm{I}}(0) = P_1(0) = \mathbf{1}$. At each internal interface between slab~$j$ and slab~$j+1$,
\begin{equation}
S_j = P_j^{-1}(\Delta x)\,\Omega_j^{-1}\,\Omega_{j+1}, 
\qquad j = 1,\dots,N-1,
\label{eq:Sjsmooth}
\end{equation}
and at the right boundary,
\begin{equation}
S_N = P_N^{-1}(\Delta x)\,\Omega_N^{-1}\,\Omega_{\mathrm{III}}.
\label{eq:SNsmooth}
\end{equation}
The total transfer matrix is
\begin{equation}
S = \prod_{j=0}^{N} S_j\,,
\label{eq:Ssmooth}
\end{equation}
from which the transmission and reflection amplitudes follow as in Eq.~\eqref{eq:tamplitudes}. In the limit $N\to\infty$ the staircase approximation converges to the exact result for the continuous profile.

\subsection*{Mode decoupling at normal incidence}
At normal incidence, the layer indices $h^{+}=-1$ and $h^{-}=+1$ appearing in $\Omega_j$ are fixed by the band index $s$ and do not depend on the local potential $V_j$. The eigenvector matrix in any slab~$j$ therefore takes the form
\begin{equation}
\Omega_j = \begin{pmatrix}
1 & 1 & 1 & 1 \\
d^{+}_{j,+} & d^{+}_{j,-} & d^{-}_{j,+} & d^{-}_{j,-} \\
-1 & -1 & 1 & 1 \\
-d^{+}_{j,+} & -d^{+}_{j,-} & d^{-}_{j,+} & d^{-}_{j,-}
\end{pmatrix},
\label{eq:Omegajky0}
\end{equation}
with $d^{s}_{j,\pm} = \pm Q_j^{s}/\bar{E}_j$.  Defining the constant transformation~\cite{barbier2010kronig,Snyman2007Ballistic}
\begin{equation}
\mathcal{T} = \begin{pmatrix}
1 & 0 & -1 & 0 \\
0 & 1 & 0 & -1 \\
1 & 0 & 1 & 0 \\
0 & 1 & 0 & 1
\end{pmatrix},
\label{eq:calT}
\end{equation}
which maps the wavefunction to the layer-symmetry basis $(\phi_{A_1}-\phi_{B_2},\,\phi_{B_1}-\phi_{A_2},\,\phi_{A_1}+\phi_{B_2},\,\phi_{B_1}+\phi_{A_2})^T$, one finds
\begin{equation}
\mathcal{T}\,\Omega_j = 2
\begin{pmatrix}
\omega_j^{(-)} & \mathbf{0} \\[2pt]
\mathbf{0} & \omega_j^{(+)}
\end{pmatrix}
\equiv 2\,\mathcal{D}_j,
\label{eq:TOmegaj}
\end{equation}
where
\begin{equation}
\omega_j^{(-)} = 
\begin{pmatrix} 
1 & 1 \\ 
d^{+}_{j,+} & d^{+}_{j,-} 
\end{pmatrix}, 
\qquad
\omega_j^{(+)} = 
\begin{pmatrix} 
1 & 1 \\ 
d^{-}_{j,+} & d^{-}_{j,-} 
\end{pmatrix}.
\label{eq:2x2blocks}
\end{equation}
Writing $\Omega_j = \mathcal{T}^{-1}\cdot 2\,\mathcal{D}_j$, the product appearing at each interface becomes
\begin{equation}
\Omega_j^{-1}\,\Omega_{j+1} 
= \mathcal{D}_j^{-1}\,\mathcal{D}_{j+1},
\label{eq:TTcancel}
\end{equation}
which is block diagonal independently of the values of $V_j$ and $V_{j+1}$. Since $P_j$ is diagonal and respects the same matrix structure, every factor $S_j$ in Eq.~\eqref{eq:Ssmooth} is block diagonal. The total transfer matrix therefore decomposes as
\begin{equation}
S = 
\begin{pmatrix} 
m^{(-)} & \mathbf{0} \\ 
\mathbf{0} & m^{(+)} 
\end{pmatrix},
\label{eq:Sblock}
\end{equation}
where $m^{(\mp)}$ are $2\times2$ matrices. As a result, no intermode conversion occurs at normal incidence: an electron incident in the $k^+$ ($k^-$) channel is transmitted and reflected exclusively within the same channel, regardless of the barrier shape. The transmission probabilities reduce to
\begin{equation}
T^{+}_{+} = \frac{1}{|m^{(-)}_{11}|^2}, 
\qquad 
T^{-}_{-} = \frac{1}{|m^{(+)}_{11}|^2},
\qquad 
T^{+}_{-} = T^{-}_{+} = 0,
\label{eq:T2x2}
\end{equation}
with $T^{s}_{s} + R^{s}_{s} = 1$ for each channel independently. This result holds for any number of slabs $N$, any set of potential values $\{V_j\}$, and therefore for any scalar potential $V(x)$ that does not introduce interlayer bias or additional coupling between the two mode sectors.

\bibliographystyle{apsrev4-1}

%


\begin{thebibliography}{61}%
\makeatletter
\providecommand \@ifxundefined [1]{%
 \@ifx{#1\undefined}
}%
\providecommand \@ifnum [1]{%
 \ifnum #1\expandafter \@firstoftwo
 \else \expandafter \@secondoftwo
 \fi
}%
\providecommand \@ifx [1]{%
 \ifx #1\expandafter \@firstoftwo
 \else \expandafter \@secondoftwo
 \fi
}%
\providecommand \natexlab [1]{#1}%
\providecommand \enquote  [1]{``#1''}%
\providecommand \bibnamefont  [1]{#1}%
\providecommand \bibfnamefont [1]{#1}%
\providecommand \citenamefont [1]{#1}%
\providecommand \href@noop [0]{\@secondoftwo}%
\providecommand \href [0]{\begingroup \@sanitize@url \@href}%
\providecommand \@href[1]{\@@startlink{#1}\@@href}%
\providecommand \@@href[1]{\endgroup#1\@@endlink}%
\providecommand \@sanitize@url [0]{\catcode `\\12\catcode `\$12\catcode `\&12\catcode `\#12\catcode `\^12\catcode `\_12\catcode `\%12\relax}%
\providecommand \@@startlink[1]{}%
\providecommand \@@endlink[0]{}%
\providecommand \url  [0]{\begingroup\@sanitize@url \@url }%
\providecommand \@url [1]{\endgroup\@href {#1}{\urlprefix }}%
\providecommand \urlprefix  [0]{URL }%
\providecommand \Eprint [0]{\href }%
\providecommand \doibase [0]{http://dx.doi.org/}%
\providecommand \selectlanguage [0]{\@gobble}%
\providecommand \bibinfo  [0]{\@secondoftwo}%
\providecommand \bibfield  [0]{\@secondoftwo}%
\providecommand \translation [1]{[#1]}%
\providecommand \BibitemOpen [0]{}%
\providecommand \bibitemStop [0]{}%
\providecommand \bibitemNoStop [0]{.\EOS\space}%
\providecommand \EOS [0]{\spacefactor3000\relax}%
\providecommand \BibitemShut  [1]{\csname bibitem#1\endcsname}%
\let\auto@bib@innerbib\@empty
\bibitem [{\citenamefont {Fallahazad}\ \emph {et~al.}(2014)\citenamefont {Fallahazad}, \citenamefont {Lee}, \citenamefont {Kang}, \citenamefont {Xue}, \citenamefont {Larentis}, \citenamefont {Corbet}, \citenamefont {Kim}, \citenamefont {Movva}, \citenamefont {Taniguchi}, \citenamefont {Watanabe}, \citenamefont {Register}, \citenamefont {Banerjee},\ and\ \citenamefont {Tutuc}}]{Fallahazad2014Gate}%
  \BibitemOpen
  \bibfield  {author} {\bibinfo {author} {\bibfnamefont {B.}~\bibnamefont {Fallahazad}}, \bibinfo {author} {\bibfnamefont {K.}~\bibnamefont {Lee}}, \bibinfo {author} {\bibfnamefont {S.}~\bibnamefont {Kang}}, \bibinfo {author} {\bibfnamefont {J.}~\bibnamefont {Xue}}, \bibinfo {author} {\bibfnamefont {S.}~\bibnamefont {Larentis}}, \bibinfo {author} {\bibfnamefont {C.}~\bibnamefont {Corbet}}, \bibinfo {author} {\bibfnamefont {K.}~\bibnamefont {Kim}}, \bibinfo {author} {\bibfnamefont {H.~C.~P.}\ \bibnamefont {Movva}}, \bibinfo {author} {\bibfnamefont {T.}~\bibnamefont {Taniguchi}}, \bibinfo {author} {\bibfnamefont {K.}~\bibnamefont {Watanabe}}, \bibinfo {author} {\bibfnamefont {L.~F.}\ \bibnamefont {Register}}, \bibinfo {author} {\bibfnamefont {S.~K.}\ \bibnamefont {Banerjee}}, \ and\ \bibinfo {author} {\bibfnamefont {E.}~\bibnamefont {Tutuc}},\ }\href {\doibase 10.1021/nl503756y} {\bibfield  {journal} {\bibinfo  {journal} {Nano Lett.}\ }\textbf {\bibinfo {volume} {15}},\ \bibinfo {pages} {428} (\bibinfo {year}
  {2014})}\BibitemShut {NoStop}%
\bibitem [{\citenamefont {Burg}\ \emph {et~al.}(2018)\citenamefont {Burg}, \citenamefont {Prasad}, \citenamefont {Kim}, \citenamefont {Taniguchi}, \citenamefont {Watanabe}, \citenamefont {MacDonald}, \citenamefont {Register},\ and\ \citenamefont {Tutuc}}]{Burg2018Strongly}%
  \BibitemOpen
  \bibfield  {author} {\bibinfo {author} {\bibfnamefont {G.~W.}\ \bibnamefont {Burg}}, \bibinfo {author} {\bibfnamefont {N.}~\bibnamefont {Prasad}}, \bibinfo {author} {\bibfnamefont {K.}~\bibnamefont {Kim}}, \bibinfo {author} {\bibfnamefont {T.}~\bibnamefont {Taniguchi}}, \bibinfo {author} {\bibfnamefont {K.}~\bibnamefont {Watanabe}}, \bibinfo {author} {\bibfnamefont {A.~H.}\ \bibnamefont {MacDonald}}, \bibinfo {author} {\bibfnamefont {L.~F.}\ \bibnamefont {Register}}, \ and\ \bibinfo {author} {\bibfnamefont {E.}~\bibnamefont {Tutuc}},\ }\href {\doibase 10.1103/PhysRevLett.120.177702} {\bibfield  {journal} {\bibinfo  {journal} {Phys. Rev. Lett.}\ }\textbf {\bibinfo {volume} {120}},\ \bibinfo {pages} {177702} (\bibinfo {year} {2018})}\BibitemShut {NoStop}%
\bibitem [{\citenamefont {Varlet}\ \emph {et~al.}(2014)\citenamefont {Varlet}, \citenamefont {Liu}, \citenamefont {Krueckl}, \citenamefont {Bischoff}, \citenamefont {Simonet}, \citenamefont {Watanabe}, \citenamefont {Taniguchi}, \citenamefont {Richter}, \citenamefont {Ensslin},\ and\ \citenamefont {Ihn}}]{Varlet2014Fabry}%
  \BibitemOpen
  \bibfield  {author} {\bibinfo {author} {\bibfnamefont {A.}~\bibnamefont {Varlet}}, \bibinfo {author} {\bibfnamefont {M.-H.}\ \bibnamefont {Liu}}, \bibinfo {author} {\bibfnamefont {V.}~\bibnamefont {Krueckl}}, \bibinfo {author} {\bibfnamefont {D.}~\bibnamefont {Bischoff}}, \bibinfo {author} {\bibfnamefont {P.}~\bibnamefont {Simonet}}, \bibinfo {author} {\bibfnamefont {K.}~\bibnamefont {Watanabe}}, \bibinfo {author} {\bibfnamefont {T.}~\bibnamefont {Taniguchi}}, \bibinfo {author} {\bibfnamefont {K.}~\bibnamefont {Richter}}, \bibinfo {author} {\bibfnamefont {K.}~\bibnamefont {Ensslin}}, \ and\ \bibinfo {author} {\bibfnamefont {T.}~\bibnamefont {Ihn}},\ }\href {\doibase 10.1103/PhysRevLett.113.116601} {\bibfield  {journal} {\bibinfo  {journal} {Phys. Rev. Lett.}\ }\textbf {\bibinfo {volume} {113}},\ \bibinfo {pages} {116601} (\bibinfo {year} {2014})}\BibitemShut {NoStop}%
\bibitem [{\citenamefont {Elahi}\ \emph {et~al.}(2024)\citenamefont {Elahi}, \citenamefont {Vakili}, \citenamefont {Zeng}, \citenamefont {Dean},\ and\ \citenamefont {Ghosh}}]{Elahi2024Direct}%
  \BibitemOpen
  \bibfield  {author} {\bibinfo {author} {\bibfnamefont {M.~M.}\ \bibnamefont {Elahi}}, \bibinfo {author} {\bibfnamefont {H.}~\bibnamefont {Vakili}}, \bibinfo {author} {\bibfnamefont {Y.}~\bibnamefont {Zeng}}, \bibinfo {author} {\bibfnamefont {C.~R.}\ \bibnamefont {Dean}}, \ and\ \bibinfo {author} {\bibfnamefont {A.~W.}\ \bibnamefont {Ghosh}},\ }\href {\doibase 10.1103/PhysRevLett.132.146302} {\bibfield  {journal} {\bibinfo  {journal} {Phys. Rev. Lett.}\ }\textbf {\bibinfo {volume} {132}},\ \bibinfo {pages} {146302} (\bibinfo {year} {2024})}\BibitemShut {NoStop}%
\bibitem [{\citenamefont {Gayduchenko}\ \emph {et~al.}(2021)\citenamefont {Gayduchenko}, \citenamefont {Xu}, \citenamefont {Alymov}, \citenamefont {Moskotin}, \citenamefont {Tretyakov}, \citenamefont {Taniguchi}, \citenamefont {Watanabe}, \citenamefont {Goltsman}, \citenamefont {Geim}, \citenamefont {Fedorov}, \citenamefont {Svintsov},\ and\ \citenamefont {Bandurin}}]{Gayduchenko2021Tunnel}%
  \BibitemOpen
  \bibfield  {author} {\bibinfo {author} {\bibfnamefont {I.}~\bibnamefont {Gayduchenko}}, \bibinfo {author} {\bibfnamefont {S.~G.}\ \bibnamefont {Xu}}, \bibinfo {author} {\bibfnamefont {G.}~\bibnamefont {Alymov}}, \bibinfo {author} {\bibfnamefont {M.}~\bibnamefont {Moskotin}}, \bibinfo {author} {\bibfnamefont {I.}~\bibnamefont {Tretyakov}}, \bibinfo {author} {\bibfnamefont {T.}~\bibnamefont {Taniguchi}}, \bibinfo {author} {\bibfnamefont {K.}~\bibnamefont {Watanabe}}, \bibinfo {author} {\bibfnamefont {G.}~\bibnamefont {Goltsman}}, \bibinfo {author} {\bibfnamefont {A.~K.}\ \bibnamefont {Geim}}, \bibinfo {author} {\bibfnamefont {G.}~\bibnamefont {Fedorov}}, \bibinfo {author} {\bibfnamefont {D.}~\bibnamefont {Svintsov}}, \ and\ \bibinfo {author} {\bibfnamefont {D.~A.}\ \bibnamefont {Bandurin}},\ }\href {\doibase 10.1038/s41467-020-20721-z} {\bibfield  {journal} {\bibinfo  {journal} {Nat. Commun.}\ }\textbf {\bibinfo {volume} {12}},\ \bibinfo {pages} {543} (\bibinfo {year} {2021})}\BibitemShut {NoStop}%
\bibitem [{\citenamefont {Beenakker}(2008)}]{Beenaker2008Colloquium}%
  \BibitemOpen
  \bibfield  {author} {\bibinfo {author} {\bibfnamefont {C.~W.~J.}\ \bibnamefont {Beenakker}},\ }\href {\doibase 10.1103/RevModPhys.80.1337} {\bibfield  {journal} {\bibinfo  {journal} {Rev. Mod. Phys.}\ }\textbf {\bibinfo {volume} {80}},\ \bibinfo {pages} {1337} (\bibinfo {year} {2008})}\BibitemShut {NoStop}%
\bibitem [{\citenamefont {Klein}(1929)}]{Klein1929Die}%
  \BibitemOpen
  \bibfield  {author} {\bibinfo {author} {\bibfnamefont {O.}~\bibnamefont {Klein}},\ }\href {\doibase 10.1007/bf01339716} {\bibfield  {journal} {\bibinfo  {journal} {Zeitschrift fur Physik}\ }\textbf {\bibinfo {volume} {53}},\ \bibinfo {pages} {157–165} (\bibinfo {year} {1929})}\BibitemShut {NoStop}%
\bibitem [{\citenamefont {Katsnelson}\ \emph {et~al.}(2006)\citenamefont {Katsnelson}, \citenamefont {Novoselov},\ and\ \citenamefont {Geim}}]{Katsnelson2006Chiral}%
  \BibitemOpen
  \bibfield  {author} {\bibinfo {author} {\bibfnamefont {M.~I.}\ \bibnamefont {Katsnelson}}, \bibinfo {author} {\bibfnamefont {K.~S.}\ \bibnamefont {Novoselov}}, \ and\ \bibinfo {author} {\bibfnamefont {A.~K.}\ \bibnamefont {Geim}},\ }\href {\doibase 10.1038/nphys384} {\bibfield  {journal} {\bibinfo  {journal} {Nat. Phys.}\ }\textbf {\bibinfo {volume} {2}},\ \bibinfo {pages} {620} (\bibinfo {year} {2006})}\BibitemShut {NoStop}%
\bibitem [{\citenamefont {Pereira}\ \emph {et~al.}(2010)\citenamefont {Pereira}, \citenamefont {Peeters}, \citenamefont {Chaves},\ and\ \citenamefont {Farias}}]{Pereira2010Klein}%
  \BibitemOpen
  \bibfield  {author} {\bibinfo {author} {\bibfnamefont {J.~M.}\ \bibnamefont {Pereira}}, \bibinfo {author} {\bibfnamefont {F.~M.}\ \bibnamefont {Peeters}}, \bibinfo {author} {\bibfnamefont {A.}~\bibnamefont {Chaves}}, \ and\ \bibinfo {author} {\bibfnamefont {G.~A.}\ \bibnamefont {Farias}},\ }\href {\doibase 10.1088/0268-1242/25/3/033002} {\bibfield  {journal} {\bibinfo  {journal} {Semicond. Sci. Technol.}\ }\textbf {\bibinfo {volume} {25}},\ \bibinfo {pages} {033002} (\bibinfo {year} {2010})}\BibitemShut {NoStop}%
\bibitem [{\citenamefont {McCann}\ and\ \citenamefont {Fal’ko}(2006)}]{mccann2006landau}%
  \BibitemOpen
  \bibfield  {author} {\bibinfo {author} {\bibfnamefont {E.}~\bibnamefont {McCann}}\ and\ \bibinfo {author} {\bibfnamefont {V.~I.}\ \bibnamefont {Fal’ko}},\ }\href {http://dx.doi.org/10.1103/PhysRevLett.96.086805} {\bibfield  {journal} {\bibinfo  {journal} {Phys. Rev. Lett.}\ }\textbf {\bibinfo {volume} {96}},\ \bibinfo {pages} {086805} (\bibinfo {year} {2006})}\BibitemShut {NoStop}%
\bibitem [{\citenamefont {Castro}\ \emph {et~al.}(2007)\citenamefont {Castro}, \citenamefont {Novoselov}, \citenamefont {Morozov}, \citenamefont {Peres}, \citenamefont {Dos~Santos}, \citenamefont {Nilsson}, \citenamefont {Guinea}, \citenamefont {Geim},\ and\ \citenamefont {Neto}}]{castro2007biased}%
  \BibitemOpen
  \bibfield  {author} {\bibinfo {author} {\bibfnamefont {E.~V.}\ \bibnamefont {Castro}}, \bibinfo {author} {\bibfnamefont {K.}~\bibnamefont {Novoselov}}, \bibinfo {author} {\bibfnamefont {S.}~\bibnamefont {Morozov}}, \bibinfo {author} {\bibfnamefont {N.}~\bibnamefont {Peres}}, \bibinfo {author} {\bibfnamefont {J.~L.}\ \bibnamefont {Dos~Santos}}, \bibinfo {author} {\bibfnamefont {J.}~\bibnamefont {Nilsson}}, \bibinfo {author} {\bibfnamefont {F.}~\bibnamefont {Guinea}}, \bibinfo {author} {\bibfnamefont {A.}~\bibnamefont {Geim}}, \ and\ \bibinfo {author} {\bibfnamefont {A.~C.}\ \bibnamefont {Neto}},\ }\href {http://dx.doi.org/10.1103/PhysRevLett.99.216802} {\bibfield  {journal} {\bibinfo  {journal} {Phys. Rev. Lett.}\ }\textbf {\bibinfo {volume} {99}},\ \bibinfo {pages} {216802} (\bibinfo {year} {2007})}\BibitemShut {NoStop}%
\bibitem [{\citenamefont {Varlet}\ \emph {et~al.}(2015)\citenamefont {Varlet}, \citenamefont {Liu}, \citenamefont {Bischoff}, \citenamefont {Simonet}, \citenamefont {Taniguchi}, \citenamefont {Watanabe}, \citenamefont {Richter}, \citenamefont {Ihn},\ and\ \citenamefont {Ensslin}}]{Varlet2015Band}%
  \BibitemOpen
  \bibfield  {author} {\bibinfo {author} {\bibfnamefont {A.}~\bibnamefont {Varlet}}, \bibinfo {author} {\bibfnamefont {M.}~\bibnamefont {Liu}}, \bibinfo {author} {\bibfnamefont {D.}~\bibnamefont {Bischoff}}, \bibinfo {author} {\bibfnamefont {P.}~\bibnamefont {Simonet}}, \bibinfo {author} {\bibfnamefont {T.}~\bibnamefont {Taniguchi}}, \bibinfo {author} {\bibfnamefont {K.}~\bibnamefont {Watanabe}}, \bibinfo {author} {\bibfnamefont {K.}~\bibnamefont {Richter}}, \bibinfo {author} {\bibfnamefont {T.}~\bibnamefont {Ihn}}, \ and\ \bibinfo {author} {\bibfnamefont {K.}~\bibnamefont {Ensslin}},\ }\href {\doibase 10.1002/pssr.201510180} {\bibfield  {journal} {\bibinfo  {journal} {Phys. Status Solidi RRL}\ }\textbf {\bibinfo {volume} {10}},\ \bibinfo {pages} {46–57} (\bibinfo {year} {2015})}\BibitemShut {NoStop}%
\bibitem [{\citenamefont {Gu}\ \emph {et~al.}(2011)\citenamefont {Gu}, \citenamefont {Rudner},\ and\ \citenamefont {Levitov}}]{Gu2011Chirality}%
  \BibitemOpen
  \bibfield  {author} {\bibinfo {author} {\bibfnamefont {N.}~\bibnamefont {Gu}}, \bibinfo {author} {\bibfnamefont {M.}~\bibnamefont {Rudner}}, \ and\ \bibinfo {author} {\bibfnamefont {L.}~\bibnamefont {Levitov}},\ }\href {\doibase 10.1103/PhysRevLett.107.156603} {\bibfield  {journal} {\bibinfo  {journal} {Phys. Rev. Lett.}\ }\textbf {\bibinfo {volume} {107}},\ \bibinfo {pages} {156603} (\bibinfo {year} {2011})}\BibitemShut {NoStop}%
\bibitem [{\citenamefont {Yamamoto}\ \emph {et~al.}(1989)\citenamefont {Yamamoto}, \citenamefont {Kanie},\ and\ \citenamefont {Taniguchi}}]{Yamamoto1989Transmission}%
  \BibitemOpen
  \bibfield  {author} {\bibinfo {author} {\bibfnamefont {H.}~\bibnamefont {Yamamoto}}, \bibinfo {author} {\bibfnamefont {Y.}~\bibnamefont {Kanie}}, \ and\ \bibinfo {author} {\bibfnamefont {K.}~\bibnamefont {Taniguchi}},\ }\href {\doibase 10.1002/pssb.2221540117} {\bibfield  {journal} {\bibinfo  {journal} {Phys. Stat. Sol. (b)}\ }\textbf {\bibinfo {volume} {154}},\ \bibinfo {pages} {195} (\bibinfo {year} {1989})}\BibitemShut {NoStop}%
\bibitem [{\citenamefont {Wu}\ \emph {et~al.}(1991)\citenamefont {Wu}, \citenamefont {Sprung}, \citenamefont {Martorell},\ and\ \citenamefont {Klarsfeld}}]{Wu1991Quantum}%
  \BibitemOpen
  \bibfield  {author} {\bibinfo {author} {\bibfnamefont {H.}~\bibnamefont {Wu}}, \bibinfo {author} {\bibfnamefont {D.~W.~L.}\ \bibnamefont {Sprung}}, \bibinfo {author} {\bibfnamefont {J.}~\bibnamefont {Martorell}}, \ and\ \bibinfo {author} {\bibfnamefont {S.}~\bibnamefont {Klarsfeld}},\ }\href {\doibase 10.1103/PhysRevB.44.6351} {\bibfield  {journal} {\bibinfo  {journal} {Phys. Rev. B}\ }\textbf {\bibinfo {volume} {44}},\ \bibinfo {pages} {6351} (\bibinfo {year} {1991})}\BibitemShut {NoStop}%
\bibitem [{\citenamefont {Ulloa}\ \emph {et~al.}(1990)\citenamefont {Ulloa}, \citenamefont {Castao},\ and\ \citenamefont {Kirczenow}}]{Ulloa1990Ballistic}%
  \BibitemOpen
  \bibfield  {author} {\bibinfo {author} {\bibfnamefont {S.~E.}\ \bibnamefont {Ulloa}}, \bibinfo {author} {\bibfnamefont {E.}~\bibnamefont {Castao}}, \ and\ \bibinfo {author} {\bibfnamefont {G.}~\bibnamefont {Kirczenow}},\ }\href {\doibase 10.1103/PhysRevB.41.12350} {\bibfield  {journal} {\bibinfo  {journal} {Phys. Rev. B}\ }\textbf {\bibinfo {volume} {41}},\ \bibinfo {pages} {12350} (\bibinfo {year} {1990})}\BibitemShut {NoStop}%
\bibitem [{\citenamefont {Nilsson}\ \emph {et~al.}(2007)\citenamefont {Nilsson}, \citenamefont {Castro~Neto}, \citenamefont {Guinea},\ and\ \citenamefont {Peres}}]{Nilsson2007Transmission}%
  \BibitemOpen
  \bibfield  {author} {\bibinfo {author} {\bibfnamefont {J.}~\bibnamefont {Nilsson}}, \bibinfo {author} {\bibfnamefont {A.~H.}\ \bibnamefont {Castro~Neto}}, \bibinfo {author} {\bibfnamefont {F.}~\bibnamefont {Guinea}}, \ and\ \bibinfo {author} {\bibfnamefont {N.~M.~R.}\ \bibnamefont {Peres}},\ }\href {\doibase 10.1103/PhysRevB.76.165416} {\bibfield  {journal} {\bibinfo  {journal} {Phys. Rev. B}\ }\textbf {\bibinfo {volume} {76}},\ \bibinfo {pages} {165416} (\bibinfo {year} {2007})}\BibitemShut {NoStop}%
\bibitem [{\citenamefont {Van~Duppen}\ and\ \citenamefont {Peeters}(2013{\natexlab{a}})}]{van2013four}%
  \BibitemOpen
  \bibfield  {author} {\bibinfo {author} {\bibfnamefont {B.}~\bibnamefont {Van~Duppen}}\ and\ \bibinfo {author} {\bibfnamefont {F.~M.}\ \bibnamefont {Peeters}},\ }\href {http://dx.doi.org/10.1103/PhysRevB.87.205427} {\bibfield  {journal} {\bibinfo  {journal} {Phys. Rev. B}\ }\textbf {\bibinfo {volume} {87}},\ \bibinfo {pages} {205427} (\bibinfo {year} {2013}{\natexlab{a}})}\BibitemShut {NoStop}%
\bibitem [{\citenamefont {Huang}\ and\ \citenamefont {Zeng}(2025)}]{Huang2025Evanescent}%
  \BibitemOpen
  \bibfield  {author} {\bibinfo {author} {\bibfnamefont {Y.}~\bibnamefont {Huang}}\ and\ \bibinfo {author} {\bibfnamefont {W.}~\bibnamefont {Zeng}},\ }\href {\doibase 10.48550/arXiv.2509.23096} {\bibfield  {journal} {\bibinfo  {journal} {arXiv}\ } (\bibinfo {year} {2025}),\ 10.48550/arXiv.2509.23096}\BibitemShut {NoStop}%
\bibitem [{\citenamefont {Agrawal~(Garg)}\ \emph {et~al.}(2012)\citenamefont {Agrawal~(Garg)}, \citenamefont {Grover}, \citenamefont {Ghosh},\ and\ \citenamefont {Sharma}}]{AgrawalGarg2012Reversal}%
  \BibitemOpen
  \bibfield  {author} {\bibinfo {author} {\bibfnamefont {N.}~\bibnamefont {Agrawal~(Garg)}}, \bibinfo {author} {\bibfnamefont {S.}~\bibnamefont {Grover}}, \bibinfo {author} {\bibfnamefont {S.}~\bibnamefont {Ghosh}}, \ and\ \bibinfo {author} {\bibfnamefont {M.}~\bibnamefont {Sharma}},\ }\href {\doibase 10.1088/0953-8984/24/17/175003} {\bibfield  {journal} {\bibinfo  {journal} {J. Phys.: Condens. Matter}\ }\textbf {\bibinfo {volume} {24}},\ \bibinfo {pages} {175003} (\bibinfo {year} {2012})}\BibitemShut {NoStop}%
\bibitem [{\citenamefont {Chen}\ and\ \citenamefont {Tao}(2009)}]{Chen2009Design}%
  \BibitemOpen
  \bibfield  {author} {\bibinfo {author} {\bibfnamefont {X.}~\bibnamefont {Chen}}\ and\ \bibinfo {author} {\bibfnamefont {J.-W.}\ \bibnamefont {Tao}},\ }\href {\doibase 10.1063/1.3168527} {\bibfield  {journal} {\bibinfo  {journal} {Appl. Phys. Lett}\ }\textbf {\bibinfo {volume} {94}},\ \bibinfo {pages} {262102} (\bibinfo {year} {2009})}\BibitemShut {NoStop}%
\bibitem [{\citenamefont {Snyman}\ and\ \citenamefont {Beenakker}(2007)}]{Snyman2007Ballistic}%
  \BibitemOpen
  \bibfield  {author} {\bibinfo {author} {\bibfnamefont {I.}~\bibnamefont {Snyman}}\ and\ \bibinfo {author} {\bibfnamefont {C.~W.~J.}\ \bibnamefont {Beenakker}},\ }\href {\doibase 10.1103/PhysRevB.75.045322} {\bibfield  {journal} {\bibinfo  {journal} {Phys. Rev. B}\ }\textbf {\bibinfo {volume} {75}},\ \bibinfo {pages} {045322} (\bibinfo {year} {2007})}\BibitemShut {NoStop}%
\bibitem [{\citenamefont {Barbier}\ \emph {et~al.}(2009)\citenamefont {Barbier}, \citenamefont {Vasilopoulos}, \citenamefont {Peeters},\ and\ \citenamefont {Pereira}}]{barbier2009bilayer}%
  \BibitemOpen
  \bibfield  {author} {\bibinfo {author} {\bibfnamefont {M.}~\bibnamefont {Barbier}}, \bibinfo {author} {\bibfnamefont {P.}~\bibnamefont {Vasilopoulos}}, \bibinfo {author} {\bibfnamefont {F.~M.}\ \bibnamefont {Peeters}}, \ and\ \bibinfo {author} {\bibfnamefont {J.~M.}\ \bibnamefont {Pereira}},\ }\href {\doibase 10.1103/physrevb.79.155402} {\bibfield  {journal} {\bibinfo  {journal} {Phys. Rev. B}\ }\textbf {\bibinfo {volume} {79}},\ \bibinfo {pages} {155402} (\bibinfo {year} {2009})}\BibitemShut {NoStop}%
\bibitem [{\citenamefont {Lee}\ \emph {et~al.}(2016)\citenamefont {Lee}, \citenamefont {Lee}, \citenamefont {Eo}, \citenamefont {Kurdak},\ and\ \citenamefont {Zhong}}]{Lee2016Evidence}%
  \BibitemOpen
  \bibfield  {author} {\bibinfo {author} {\bibfnamefont {K.}~\bibnamefont {Lee}}, \bibinfo {author} {\bibfnamefont {S.}~\bibnamefont {Lee}}, \bibinfo {author} {\bibfnamefont {Y.~S.}\ \bibnamefont {Eo}}, \bibinfo {author} {\bibfnamefont {C.}~\bibnamefont {Kurdak}}, \ and\ \bibinfo {author} {\bibfnamefont {Z.}~\bibnamefont {Zhong}},\ }\href {\doibase 10.1103/PhysRevB.94.205418} {\bibfield  {journal} {\bibinfo  {journal} {Phys. Rev. B}\ }\textbf {\bibinfo {volume} {94}},\ \bibinfo {pages} {205418} (\bibinfo {year} {2016})}\BibitemShut {NoStop}%
\bibitem [{\citenamefont {Park}\ and\ \citenamefont {Sim}(2011)}]{Park2011Pi}%
  \BibitemOpen
  \bibfield  {author} {\bibinfo {author} {\bibfnamefont {S.}~\bibnamefont {Park}}\ and\ \bibinfo {author} {\bibfnamefont {H.-S.}\ \bibnamefont {Sim}},\ }\href {\doibase 10.1103/PhysRevB.84.235432} {\bibfield  {journal} {\bibinfo  {journal} {Phys. Rev. B}\ }\textbf {\bibinfo {volume} {84}},\ \bibinfo {pages} {235432} (\bibinfo {year} {2011})}\BibitemShut {NoStop}%
\bibitem [{\citenamefont {Lu}\ \emph {et~al.}(2015)\citenamefont {Lu}, \citenamefont {Li}, \citenamefont {Xu},\ and\ \citenamefont {Ye}}]{Lu2015Destruction}%
  \BibitemOpen
  \bibfield  {author} {\bibinfo {author} {\bibfnamefont {W.}~\bibnamefont {Lu}}, \bibinfo {author} {\bibfnamefont {W.}~\bibnamefont {Li}}, \bibinfo {author} {\bibfnamefont {C.}~\bibnamefont {Xu}}, \ and\ \bibinfo {author} {\bibfnamefont {C.}~\bibnamefont {Ye}},\ }\href {\doibase 10.1088/0022-3727/48/28/285102} {\bibfield  {journal} {\bibinfo  {journal} {J. Phys. D: Appl. Phys.}\ }\textbf {\bibinfo {volume} {48}},\ \bibinfo {pages} {285102} (\bibinfo {year} {2015})}\BibitemShut {NoStop}%
\bibitem [{\citenamefont {Lamas-Mart\'{\i}nez}\ \emph {et~al.}(2024)\citenamefont {Lamas-Mart\'{\i}nez}, \citenamefont {Briones-Torres}, \citenamefont {Molina-Valdovinos},\ and\ \citenamefont {Rodr\'{\i}guez-Vargas}}]{Lamas2024Persistent}%
  \BibitemOpen
  \bibfield  {author} {\bibinfo {author} {\bibfnamefont {K.~J.}\ \bibnamefont {Lamas-Mart\'{\i}nez}}, \bibinfo {author} {\bibfnamefont {J.~A.}\ \bibnamefont {Briones-Torres}}, \bibinfo {author} {\bibfnamefont {S.}~\bibnamefont {Molina-Valdovinos}}, \ and\ \bibinfo {author} {\bibfnamefont {I.}~\bibnamefont {Rodr\'{\i}guez-Vargas}},\ }\href {\doibase 10.1103/PhysRevB.109.035416} {\bibfield  {journal} {\bibinfo  {journal} {Phys. Rev. B}\ }\textbf {\bibinfo {volume} {109}},\ \bibinfo {pages} {035416} (\bibinfo {year} {2024})}\BibitemShut {NoStop}%
\bibitem [{\citenamefont {Zhao}\ and\ \citenamefont {Chen}(2011)}]{Zhao2011Electronic}%
  \BibitemOpen
  \bibfield  {author} {\bibinfo {author} {\bibfnamefont {P.-L.}\ \bibnamefont {Zhao}}\ and\ \bibinfo {author} {\bibfnamefont {X.}~\bibnamefont {Chen}},\ }\href {\doibase 10.1063/1.3658394} {\bibfield  {journal} {\bibinfo  {journal} {Appl. Phys. Lett}\ }\textbf {\bibinfo {volume} {99}},\ \bibinfo {pages} {182108} (\bibinfo {year} {2011})}\BibitemShut {NoStop}%
\bibitem [{\citenamefont {Wang}\ and\ \citenamefont {Chen}(2011)}]{Wang2011Robust}%
  \BibitemOpen
  \bibfield  {author} {\bibinfo {author} {\bibfnamefont {L.-G.}\ \bibnamefont {Wang}}\ and\ \bibinfo {author} {\bibfnamefont {X.}~\bibnamefont {Chen}},\ }\href {\doibase 10.1063/1.3525270} {\bibfield  {journal} {\bibinfo  {journal} {J. Appl. Phys.}\ }\textbf {\bibinfo {volume} {109}},\ \bibinfo {pages} {033710} (\bibinfo {year} {2011})}\BibitemShut {NoStop}%
\bibitem [{\citenamefont {Jing}\ \emph {et~al.}(2010)\citenamefont {Jing}, \citenamefont {Velasco~Jr.}, \citenamefont {Kratz}, \citenamefont {Liu}, \citenamefont {Bao}, \citenamefont {Bockrath},\ and\ \citenamefont {Lau}}]{Jing2010Quantum}%
  \BibitemOpen
  \bibfield  {author} {\bibinfo {author} {\bibfnamefont {L.}~\bibnamefont {Jing}}, \bibinfo {author} {\bibfnamefont {J.}~\bibnamefont {Velasco~Jr.}}, \bibinfo {author} {\bibfnamefont {P.}~\bibnamefont {Kratz}}, \bibinfo {author} {\bibfnamefont {G.}~\bibnamefont {Liu}}, \bibinfo {author} {\bibfnamefont {W.}~\bibnamefont {Bao}}, \bibinfo {author} {\bibfnamefont {M.}~\bibnamefont {Bockrath}}, \ and\ \bibinfo {author} {\bibfnamefont {C.~N.}\ \bibnamefont {Lau}},\ }\href {\doibase 10.1021/nl101901g} {\bibfield  {journal} {\bibinfo  {journal} {Nano Lett.}\ }\textbf {\bibinfo {volume} {10}},\ \bibinfo {pages} {4000} (\bibinfo {year} {2010})}\BibitemShut {NoStop}%
\bibitem [{\citenamefont {Mayorov}\ \emph {et~al.}(2011)\citenamefont {Mayorov}, \citenamefont {Elias}, \citenamefont {Mucha-Kruczynski}, \citenamefont {Gorbachev}, \citenamefont {Tudorovskiy}, \citenamefont {Zhukov}, \citenamefont {Morozov}, \citenamefont {Katsnelson}, \citenamefont {Geim},\ and\ \citenamefont {Novoselov}}]{Mayorov2011Interaction}%
  \BibitemOpen
  \bibfield  {author} {\bibinfo {author} {\bibfnamefont {A.~S.}\ \bibnamefont {Mayorov}}, \bibinfo {author} {\bibfnamefont {D.~C.}\ \bibnamefont {Elias}}, \bibinfo {author} {\bibfnamefont {M.}~\bibnamefont {Mucha-Kruczynski}}, \bibinfo {author} {\bibfnamefont {R.~V.}\ \bibnamefont {Gorbachev}}, \bibinfo {author} {\bibfnamefont {T.}~\bibnamefont {Tudorovskiy}}, \bibinfo {author} {\bibfnamefont {A.}~\bibnamefont {Zhukov}}, \bibinfo {author} {\bibfnamefont {S.~V.}\ \bibnamefont {Morozov}}, \bibinfo {author} {\bibfnamefont {M.~I.}\ \bibnamefont {Katsnelson}}, \bibinfo {author} {\bibfnamefont {A.~K.}\ \bibnamefont {Geim}}, \ and\ \bibinfo {author} {\bibfnamefont {K.~S.}\ \bibnamefont {Novoselov}},\ }\href {\doibase 10.1126/science.1208683} {\bibfield  {journal} {\bibinfo  {journal} {Science}\ }\textbf {\bibinfo {volume} {333}},\ \bibinfo {pages} {860} (\bibinfo {year} {2011})}\BibitemShut {NoStop}%
\bibitem [{\citenamefont {Kleptsyn}\ \emph {et~al.}(2015)\citenamefont {Kleptsyn}, \citenamefont {Okunev}, \citenamefont {Schurov}, \citenamefont {Zubov},\ and\ \citenamefont {Katsnelson}}]{Kleptsyn2015Chiral}%
  \BibitemOpen
  \bibfield  {author} {\bibinfo {author} {\bibfnamefont {V.}~\bibnamefont {Kleptsyn}}, \bibinfo {author} {\bibfnamefont {A.}~\bibnamefont {Okunev}}, \bibinfo {author} {\bibfnamefont {I.}~\bibnamefont {Schurov}}, \bibinfo {author} {\bibfnamefont {D.}~\bibnamefont {Zubov}}, \ and\ \bibinfo {author} {\bibfnamefont {M.~I.}\ \bibnamefont {Katsnelson}},\ }\href {\doibase 10.1103/physrevb.92.165407} {\bibfield  {journal} {\bibinfo  {journal} {Phys. Rev. B}\ }\textbf {\bibinfo {volume} {92}},\ \bibinfo {pages} {165407} (\bibinfo {year} {2015})}\BibitemShut {NoStop}%
\bibitem [{\citenamefont {Ando}(1991)}]{Ando1991Quantum}%
  \BibitemOpen
  \bibfield  {author} {\bibinfo {author} {\bibfnamefont {T.}~\bibnamefont {Ando}},\ }\href {\doibase 10.1103/PhysRevB.44.8017} {\bibfield  {journal} {\bibinfo  {journal} {Phys. Rev. B}\ }\textbf {\bibinfo {volume} {44}},\ \bibinfo {pages} {8017} (\bibinfo {year} {1991})}\BibitemShut {NoStop}%
\bibitem [{\citenamefont {Sanderson}\ \emph {et~al.}(2013)\citenamefont {Sanderson}, \citenamefont {Ang},\ and\ \citenamefont {Zhang}}]{Sanderson2013Klein}%
  \BibitemOpen
  \bibfield  {author} {\bibinfo {author} {\bibfnamefont {M.}~\bibnamefont {Sanderson}}, \bibinfo {author} {\bibfnamefont {Y.~S.}\ \bibnamefont {Ang}}, \ and\ \bibinfo {author} {\bibfnamefont {C.}~\bibnamefont {Zhang}},\ }\href {\doibase 10.1103/PhysRevB.88.245404} {\bibfield  {journal} {\bibinfo  {journal} {Phys. Rev. B}\ }\textbf {\bibinfo {volume} {88}},\ \bibinfo {pages} {245404} (\bibinfo {year} {2013})}\BibitemShut {NoStop}%
\bibitem [{\citenamefont {Hassane~Saley}\ \emph {et~al.}(2022)\citenamefont {Hassane~Saley}, \citenamefont {El~Mouhafid}, \citenamefont {Jellal},\ and\ \citenamefont {Siari}}]{HassaneSaley2022Klein}%
  \BibitemOpen
  \bibfield  {author} {\bibinfo {author} {\bibfnamefont {M.}~\bibnamefont {Hassane~Saley}}, \bibinfo {author} {\bibfnamefont {A.}~\bibnamefont {El~Mouhafid}}, \bibinfo {author} {\bibfnamefont {A.}~\bibnamefont {Jellal}}, \ and\ \bibinfo {author} {\bibfnamefont {A.}~\bibnamefont {Siari}},\ }\href {\doibase 10.1002/andp.202200308} {\bibfield  {journal} {\bibinfo  {journal} {Ann. Phys.}\ }\textbf {\bibinfo {volume} {534}},\ \bibinfo {pages} {2200308} (\bibinfo {year} {2022})}\BibitemShut {NoStop}%
\bibitem [{\citenamefont {Dell’Anna}\ \emph {et~al.}(2018)\citenamefont {Dell’Anna}, \citenamefont {Majari},\ and\ \citenamefont {Setare}}]{DellAnna2018From}%
  \BibitemOpen
  \bibfield  {author} {\bibinfo {author} {\bibfnamefont {L.}~\bibnamefont {Dell’Anna}}, \bibinfo {author} {\bibfnamefont {P.}~\bibnamefont {Majari}}, \ and\ \bibinfo {author} {\bibfnamefont {M.~R.}\ \bibnamefont {Setare}},\ }\href {\doibase 10.1088/1361-648x/aadf2e} {\bibfield  {journal} {\bibinfo  {journal} {J. Phys.: Condens. Matter}\ }\textbf {\bibinfo {volume} {30}},\ \bibinfo {pages} {415301} (\bibinfo {year} {2018})}\BibitemShut {NoStop}%
\bibitem [{\citenamefont {Van~Duppen}\ \emph {et~al.}(2013)\citenamefont {Van~Duppen}, \citenamefont {Sena},\ and\ \citenamefont {Peeters}}]{Van2013Multiband}%
  \BibitemOpen
  \bibfield  {author} {\bibinfo {author} {\bibfnamefont {B.}~\bibnamefont {Van~Duppen}}, \bibinfo {author} {\bibfnamefont {S.~H.~R.}\ \bibnamefont {Sena}}, \ and\ \bibinfo {author} {\bibfnamefont {F.~M.}\ \bibnamefont {Peeters}},\ }\href {\doibase 10.1103/PhysRevB.87.195439} {\bibfield  {journal} {\bibinfo  {journal} {Phys. Rev. B}\ }\textbf {\bibinfo {volume} {87}},\ \bibinfo {pages} {195439} (\bibinfo {year} {2013})}\BibitemShut {NoStop}%
\bibitem [{\citenamefont {García}\ \emph {et~al.}(2022)\citenamefont {García}, \citenamefont {Stegmann},\ and\ \citenamefont {Betancur-Ocampo}}]{Garca2022Generalized}%
  \BibitemOpen
  \bibfield  {author} {\bibinfo {author} {\bibfnamefont {S.~G.~y.}\ \bibnamefont {García}}, \bibinfo {author} {\bibfnamefont {T.}~\bibnamefont {Stegmann}}, \ and\ \bibinfo {author} {\bibfnamefont {Y.}~\bibnamefont {Betancur-Ocampo}},\ }\href {\doibase 10.1103/physrevb.105.125139} {\bibfield  {journal} {\bibinfo  {journal} {Phys. Rev. B}\ }\textbf {\bibinfo {volume} {105}},\ \bibinfo {pages} {125139} (\bibinfo {year} {2022})}\BibitemShut {NoStop}%
\bibitem [{\citenamefont {He}\ \emph {et~al.}(2013)\citenamefont {He}, \citenamefont {Chu},\ and\ \citenamefont {He}}]{He2013Chiral}%
  \BibitemOpen
  \bibfield  {author} {\bibinfo {author} {\bibfnamefont {W.-Y.}\ \bibnamefont {He}}, \bibinfo {author} {\bibfnamefont {Z.-D.}\ \bibnamefont {Chu}}, \ and\ \bibinfo {author} {\bibfnamefont {L.}~\bibnamefont {He}},\ }\href {\doibase 10.1103/PhysRevLett.111.066803} {\bibfield  {journal} {\bibinfo  {journal} {Phys. Rev. Lett.}\ }\textbf {\bibinfo {volume} {111}},\ \bibinfo {pages} {066803} (\bibinfo {year} {2013})}\BibitemShut {NoStop}%
\bibitem [{\citenamefont {Handschin}\ \emph {et~al.}(2016)\citenamefont {Handschin}, \citenamefont {Makk}, \citenamefont {Rickhaus}, \citenamefont {Liu}, \citenamefont {Watanabe}, \citenamefont {Taniguchi}, \citenamefont {Richter},\ and\ \citenamefont {Sch\"{o}nenberger}}]{Handschin2016Fabry}%
  \BibitemOpen
  \bibfield  {author} {\bibinfo {author} {\bibfnamefont {C.}~\bibnamefont {Handschin}}, \bibinfo {author} {\bibfnamefont {P.}~\bibnamefont {Makk}}, \bibinfo {author} {\bibfnamefont {P.}~\bibnamefont {Rickhaus}}, \bibinfo {author} {\bibfnamefont {M.-H.}\ \bibnamefont {Liu}}, \bibinfo {author} {\bibfnamefont {K.}~\bibnamefont {Watanabe}}, \bibinfo {author} {\bibfnamefont {T.}~\bibnamefont {Taniguchi}}, \bibinfo {author} {\bibfnamefont {K.}~\bibnamefont {Richter}}, \ and\ \bibinfo {author} {\bibfnamefont {C.}~\bibnamefont {Sch\"{o}nenberger}},\ }\href {\doibase 10.1021/acs.nanolett.6b04137} {\bibfield  {journal} {\bibinfo  {journal} {Nano Lett.}\ }\textbf {\bibinfo {volume} {17}},\ \bibinfo {pages} {328–333} (\bibinfo {year} {2016})}\BibitemShut {NoStop}%
\bibitem [{\citenamefont {Shytov}(2015)}]{Shytov2015Cloaked}%
  \BibitemOpen
  \bibfield  {author} {\bibinfo {author} {\bibfnamefont {A.~V.}\ \bibnamefont {Shytov}},\ }\href {\doibase 10.48550/arXiv.1506.02839} {\bibfield  {journal} {\bibinfo  {journal} {arXiv}\ } (\bibinfo {year} {2015}),\ 10.48550/arXiv.1506.02839}\BibitemShut {NoStop}%
\bibitem [{\citenamefont {Van~der Donck}\ \emph {et~al.}(2016)\citenamefont {Van~der Donck}, \citenamefont {Peeters},\ and\ \citenamefont {Van~Duppen}}]{Van2016Transport}%
  \BibitemOpen
  \bibfield  {author} {\bibinfo {author} {\bibfnamefont {M.}~\bibnamefont {Van~der Donck}}, \bibinfo {author} {\bibfnamefont {F.~M.}\ \bibnamefont {Peeters}}, \ and\ \bibinfo {author} {\bibfnamefont {B.}~\bibnamefont {Van~Duppen}},\ }\href {\doibase 10.1103/PhysRevB.93.115423} {\bibfield  {journal} {\bibinfo  {journal} {Phys. Rev. B}\ }\textbf {\bibinfo {volume} {93}},\ \bibinfo {pages} {115423} (\bibinfo {year} {2016})}\BibitemShut {NoStop}%
\bibitem [{\citenamefont {Hund}(1927)}]{Hund1927}%
  \BibitemOpen
  \bibfield  {author} {\bibinfo {author} {\bibfnamefont {F.}~\bibnamefont {Hund}},\ }\href {\doibase 10.1007/bf01400234} {\bibfield  {journal} {\bibinfo  {journal} {Z. Phys.}\ }\textbf {\bibinfo {volume} {40}},\ \bibinfo {pages} {742} (\bibinfo {year} {1927})}\BibitemShut {NoStop}%
\bibitem [{\citenamefont {Tsu}\ and\ \citenamefont {Esaki}(1973)}]{Tsu1973Tunneling}%
  \BibitemOpen
  \bibfield  {author} {\bibinfo {author} {\bibfnamefont {R.}~\bibnamefont {Tsu}}\ and\ \bibinfo {author} {\bibfnamefont {L.}~\bibnamefont {Esaki}},\ }\href {\doibase 10.1063/1.1654509} {\bibfield  {journal} {\bibinfo  {journal} {Appl. Phys. Lett}\ }\textbf {\bibinfo {volume} {22}},\ \bibinfo {pages} {562} (\bibinfo {year} {1973})}\BibitemShut {NoStop}%
\bibitem [{\citenamefont {Van~Duppen}\ and\ \citenamefont {Peeters}(2013{\natexlab{b}})}]{VanDuppen2013Klein}%
  \BibitemOpen
  \bibfield  {author} {\bibinfo {author} {\bibfnamefont {B.}~\bibnamefont {Van~Duppen}}\ and\ \bibinfo {author} {\bibfnamefont {F.~M.}\ \bibnamefont {Peeters}},\ }\href {\doibase 10.1209/0295-5075/102/27001} {\bibfield  {journal} {\bibinfo  {journal} {Europhys. Lett.}\ }\textbf {\bibinfo {volume} {102}},\ \bibinfo {pages} {27001} (\bibinfo {year} {2013}{\natexlab{b}})}\BibitemShut {NoStop}%
\bibitem [{\citenamefont {Dragoman}(2008)}]{Dragoman2008Evidence}%
  \BibitemOpen
  \bibfield  {author} {\bibinfo {author} {\bibfnamefont {D.}~\bibnamefont {Dragoman}},\ }\href {\doibase 10.1088/0031-8949/79/01/015003} {\bibfield  {journal} {\bibinfo  {journal} {Phys. Scr.}\ }\textbf {\bibinfo {volume} {79}},\ \bibinfo {pages} {015003} (\bibinfo {year} {2008})}\BibitemShut {NoStop}%
\bibitem [{\citenamefont {Alvarado-Goytia}\ \emph {et~al.}(2022)\citenamefont {Alvarado-Goytia}, \citenamefont {Rodríguez-González}, \citenamefont {Martínez-Orozco},\ and\ \citenamefont {Rodríguez-Vargas}}]{Alvarado2022Biperiodic}%
  \BibitemOpen
  \bibfield  {author} {\bibinfo {author} {\bibfnamefont {J.~J.}\ \bibnamefont {Alvarado-Goytia}}, \bibinfo {author} {\bibfnamefont {R.}~\bibnamefont {Rodríguez-González}}, \bibinfo {author} {\bibfnamefont {J.~C.}\ \bibnamefont {Martínez-Orozco}}, \ and\ \bibinfo {author} {\bibfnamefont {I.}~\bibnamefont {Rodríguez-Vargas}},\ }\href {\doibase 10.1038/s41598-021-04690-x} {\bibfield  {journal} {\bibinfo  {journal} {Sci. Rep.}\ }\textbf {\bibinfo {volume} {12}},\ \bibinfo {pages} {832} (\bibinfo {year} {2022})}\BibitemShut {NoStop}%
\bibitem [{\citenamefont {Iogansen}(1987)}]{Iogansen1987}%
  \BibitemOpen
  \bibfield  {author} {\bibinfo {author} {\bibfnamefont {L.~V.}\ \bibnamefont {Iogansen}},\ }\href@noop {} {\bibfield  {journal} {\bibinfo  {journal} {Soviet Technical Physics Letters}\ }\textbf {\bibinfo {volume} {13}},\ \bibinfo {pages} {478} (\bibinfo {year} {1987})}\BibitemShut {NoStop}%
\bibitem [{\citenamefont {Xu}\ \emph {et~al.}(2014)\citenamefont {Xu}, \citenamefont {He},\ and\ \citenamefont {Yang}}]{Xu2014Resonant}%
  \BibitemOpen
  \bibfield  {author} {\bibinfo {author} {\bibfnamefont {Y.}~\bibnamefont {Xu}}, \bibinfo {author} {\bibfnamefont {Y.}~\bibnamefont {He}}, \ and\ \bibinfo {author} {\bibfnamefont {Y.}~\bibnamefont {Yang}},\ }\href {\doibase 10.1007/s00339-014-8423-2} {\bibfield  {journal} {\bibinfo  {journal} {Appl. Phys. A: Mater. Sci. Process.}\ }\textbf {\bibinfo {volume} {115}},\ \bibinfo {pages} {721} (\bibinfo {year} {2014})}\BibitemShut {NoStop}%
\bibitem [{\citenamefont {Xu}\ \emph {et~al.}(2019)\citenamefont {Xu}, \citenamefont {Feng},\ and\ \citenamefont {Zhang}}]{Xu2019Resonant}%
  \BibitemOpen
  \bibfield  {author} {\bibinfo {author} {\bibfnamefont {H.~Z.}\ \bibnamefont {Xu}}, \bibinfo {author} {\bibfnamefont {S.}~\bibnamefont {Feng}}, \ and\ \bibinfo {author} {\bibfnamefont {Y.}~\bibnamefont {Zhang}},\ }\href {\doibase 10.1007/s11082-019-1873-1} {\bibfield  {journal} {\bibinfo  {journal} {Opt. Quantum Electron.}\ }\textbf {\bibinfo {volume} {51}},\ \bibinfo {pages} {158} (\bibinfo {year} {2019})}\BibitemShut {NoStop}%
\bibitem [{\citenamefont {Garc\'{\i}a-Calder\'on}\ \emph {et~al.}(1993)\citenamefont {Garc\'{\i}a-Calder\'on}, \citenamefont {Romo},\ and\ \citenamefont {Rubio}}]{Garcia1993Descripcion}%
  \BibitemOpen
  \bibfield  {author} {\bibinfo {author} {\bibfnamefont {G.}~\bibnamefont {Garc\'{\i}a-Calder\'on}}, \bibinfo {author} {\bibfnamefont {R.}~\bibnamefont {Romo}}, \ and\ \bibinfo {author} {\bibfnamefont {A.}~\bibnamefont {Rubio}},\ }\href {\doibase 10.1103/PhysRevB.47.9572} {\bibfield  {journal} {\bibinfo  {journal} {Phys. Rev. B}\ }\textbf {\bibinfo {volume} {47}},\ \bibinfo {pages} {9572} (\bibinfo {year} {1993})}\BibitemShut {NoStop}%
\bibitem [{\citenamefont {Romo}\ and\ \citenamefont {Garc\'{\i}a-Calder\'on}(1994)}]{Garcia1994Descripcion}%
  \BibitemOpen
  \bibfield  {author} {\bibinfo {author} {\bibfnamefont {R.}~\bibnamefont {Romo}}\ and\ \bibinfo {author} {\bibfnamefont {G.}~\bibnamefont {Garc\'{\i}a-Calder\'on}},\ }\href {\doibase 10.1103/PhysRevB.49.14016} {\bibfield  {journal} {\bibinfo  {journal} {Phys. Rev. B}\ }\textbf {\bibinfo {volume} {49}},\ \bibinfo {pages} {14016} (\bibinfo {year} {1994})}\BibitemShut {NoStop}%
\bibitem [{\citenamefont {Baringhaus}\ \emph {et~al.}(2015)\citenamefont {Baringhaus}, \citenamefont {St\"{o}hr}, \citenamefont {Forti}, \citenamefont {Starke},\ and\ \citenamefont {Tegenkamp}}]{Baringhaus2015Ballistic}%
  \BibitemOpen
  \bibfield  {author} {\bibinfo {author} {\bibfnamefont {J.}~\bibnamefont {Baringhaus}}, \bibinfo {author} {\bibfnamefont {A.}~\bibnamefont {St\"{o}hr}}, \bibinfo {author} {\bibfnamefont {S.}~\bibnamefont {Forti}}, \bibinfo {author} {\bibfnamefont {U.}~\bibnamefont {Starke}}, \ and\ \bibinfo {author} {\bibfnamefont {C.}~\bibnamefont {Tegenkamp}},\ }\href {\doibase 10.1038/srep09955} {\bibfield  {journal} {\bibinfo  {journal} {Scientific Reports}\ }\textbf {\bibinfo {volume} {5}} (\bibinfo {year} {2015}),\ 10.1038/srep09955}\BibitemShut {NoStop}%
\bibitem [{\citenamefont {Huard}\ \emph {et~al.}(2007)\citenamefont {Huard}, \citenamefont {Sulpizio}, \citenamefont {Stander}, \citenamefont {Todd}, \citenamefont {Yang},\ and\ \citenamefont {Goldhaber-Gordon}}]{Huard2007Transport}%
  \BibitemOpen
  \bibfield  {author} {\bibinfo {author} {\bibfnamefont {B.}~\bibnamefont {Huard}}, \bibinfo {author} {\bibfnamefont {J.~A.}\ \bibnamefont {Sulpizio}}, \bibinfo {author} {\bibfnamefont {N.}~\bibnamefont {Stander}}, \bibinfo {author} {\bibfnamefont {K.}~\bibnamefont {Todd}}, \bibinfo {author} {\bibfnamefont {B.}~\bibnamefont {Yang}}, \ and\ \bibinfo {author} {\bibfnamefont {D.}~\bibnamefont {Goldhaber-Gordon}},\ }\href {\doibase 10.1103/PhysRevLett.98.236803} {\bibfield  {journal} {\bibinfo  {journal} {Phys. Rev. Lett.}\ }\textbf {\bibinfo {volume} {98}},\ \bibinfo {pages} {236803} (\bibinfo {year} {2007})}\BibitemShut {NoStop}%
\bibitem [{\citenamefont {Campos}\ \emph {et~al.}(2012)\citenamefont {Campos}, \citenamefont {Young}, \citenamefont {Surakitbovorn}, \citenamefont {Watanabe}, \citenamefont {Taniguchi},\ and\ \citenamefont {Jarillo-Herrero}}]{Campos2012Quantum}%
  \BibitemOpen
  \bibfield  {author} {\bibinfo {author} {\bibfnamefont {L.}~\bibnamefont {Campos}}, \bibinfo {author} {\bibfnamefont {A.}~\bibnamefont {Young}}, \bibinfo {author} {\bibfnamefont {K.}~\bibnamefont {Surakitbovorn}}, \bibinfo {author} {\bibfnamefont {K.}~\bibnamefont {Watanabe}}, \bibinfo {author} {\bibfnamefont {T.}~\bibnamefont {Taniguchi}}, \ and\ \bibinfo {author} {\bibfnamefont {P.}~\bibnamefont {Jarillo-Herrero}},\ }\href {\doibase 10.1038/ncomms2243} {\bibfield  {journal} {\bibinfo  {journal} {Nat. Commun.}\ }\textbf {\bibinfo {volume} {3}},\ \bibinfo {pages} {1239} (\bibinfo {year} {2012})}\BibitemShut {NoStop}%
\bibitem [{\citenamefont {Pereira}\ and\ \citenamefont {Katsnelson}(2015)}]{Pereira2015Landau}%
  \BibitemOpen
  \bibfield  {author} {\bibinfo {author} {\bibfnamefont {J.~M.}\ \bibnamefont {Pereira}}\ and\ \bibinfo {author} {\bibfnamefont {M.~I.}\ \bibnamefont {Katsnelson}},\ }\href {\doibase 10.1103/physrevb.92.075437} {\bibfield  {journal} {\bibinfo  {journal} {Phys. Rev. B}\ }\textbf {\bibinfo {volume} {92}},\ \bibinfo {pages} {075437} (\bibinfo {year} {2015})}\BibitemShut {NoStop}%
\bibitem [{\citenamefont {Young}\ and\ \citenamefont {Kim}(2009)}]{Young2009Quantum}%
  \BibitemOpen
  \bibfield  {author} {\bibinfo {author} {\bibfnamefont {A.~F.}\ \bibnamefont {Young}}\ and\ \bibinfo {author} {\bibfnamefont {P.}~\bibnamefont {Kim}},\ }\href {\doibase 10.1038/nphys1198} {\bibfield  {journal} {\bibinfo  {journal} {Nature Phys.}\ }\textbf {\bibinfo {volume} {5}},\ \bibinfo {pages} {222–226} (\bibinfo {year} {2009})}\BibitemShut {NoStop}%
\bibitem [{\citenamefont {Stander}\ \emph {et~al.}(2009)\citenamefont {Stander}, \citenamefont {Huard},\ and\ \citenamefont {Goldhaber-Gordon}}]{Stander2009Evidence}%
  \BibitemOpen
  \bibfield  {author} {\bibinfo {author} {\bibfnamefont {N.}~\bibnamefont {Stander}}, \bibinfo {author} {\bibfnamefont {B.}~\bibnamefont {Huard}}, \ and\ \bibinfo {author} {\bibfnamefont {D.}~\bibnamefont {Goldhaber-Gordon}},\ }\href {\doibase 10.1103/physrevlett.102.026807} {\bibfield  {journal} {\bibinfo  {journal} {Phys. Rev. Lett.}\ }\textbf {\bibinfo {volume} {102}},\ \bibinfo {pages} {026807} (\bibinfo {year} {2009})}\BibitemShut {NoStop}%
\bibitem [{\citenamefont {Mylnikov}\ \emph {et~al.}(2022)\citenamefont {Mylnikov}, \citenamefont {Titova}, \citenamefont {Kashchenko}, \citenamefont {Safonov}, \citenamefont {Zhukov}, \citenamefont {Semkin}, \citenamefont {Novoselov}, \citenamefont {Bandurin},\ and\ \citenamefont {Svintsov}}]{Mylnikov2022Terahertz}%
  \BibitemOpen
  \bibfield  {author} {\bibinfo {author} {\bibfnamefont {D.~A.}\ \bibnamefont {Mylnikov}}, \bibinfo {author} {\bibfnamefont {E.~I.}\ \bibnamefont {Titova}}, \bibinfo {author} {\bibfnamefont {M.~A.}\ \bibnamefont {Kashchenko}}, \bibinfo {author} {\bibfnamefont {I.~V.}\ \bibnamefont {Safonov}}, \bibinfo {author} {\bibfnamefont {S.~S.}\ \bibnamefont {Zhukov}}, \bibinfo {author} {\bibfnamefont {V.~A.}\ \bibnamefont {Semkin}}, \bibinfo {author} {\bibfnamefont {K.~S.}\ \bibnamefont {Novoselov}}, \bibinfo {author} {\bibfnamefont {D.~A.}\ \bibnamefont {Bandurin}}, \ and\ \bibinfo {author} {\bibfnamefont {D.~A.}\ \bibnamefont {Svintsov}},\ }\href {\doibase 10.1021/acs.nanolett.2c04119} {\bibfield  {journal} {\bibinfo  {journal} {Nano Lett.}\ }\textbf {\bibinfo {volume} {23}},\ \bibinfo {pages} {220–226} (\bibinfo {year} {2022})}\BibitemShut {NoStop}%
\bibitem [{\citenamefont {Titova}\ \emph {et~al.}(2023)\citenamefont {Titova}, \citenamefont {Mylnikov}, \citenamefont {Kashchenko}, \citenamefont {Safonov}, \citenamefont {Zhukov}, \citenamefont {Dzhikirba}, \citenamefont {Novoselov}, \citenamefont {Bandurin}, \citenamefont {Alymov},\ and\ \citenamefont {Svintsov}}]{Titova2023Ultralow}%
  \BibitemOpen
  \bibfield  {author} {\bibinfo {author} {\bibfnamefont {E.}~\bibnamefont {Titova}}, \bibinfo {author} {\bibfnamefont {D.}~\bibnamefont {Mylnikov}}, \bibinfo {author} {\bibfnamefont {M.}~\bibnamefont {Kashchenko}}, \bibinfo {author} {\bibfnamefont {I.}~\bibnamefont {Safonov}}, \bibinfo {author} {\bibfnamefont {S.}~\bibnamefont {Zhukov}}, \bibinfo {author} {\bibfnamefont {K.}~\bibnamefont {Dzhikirba}}, \bibinfo {author} {\bibfnamefont {K.~S.}\ \bibnamefont {Novoselov}}, \bibinfo {author} {\bibfnamefont {D.~A.}\ \bibnamefont {Bandurin}}, \bibinfo {author} {\bibfnamefont {G.}~\bibnamefont {Alymov}}, \ and\ \bibinfo {author} {\bibfnamefont {D.}~\bibnamefont {Svintsov}},\ }\href {\doibase 10.1021/acsnano.2c12285} {\bibfield  {journal} {\bibinfo  {journal} {ACS Nano}\ }\textbf {\bibinfo {volume} {17}},\ \bibinfo {pages} {8223–8232} (\bibinfo {year} {2023})}\BibitemShut {NoStop}%
\bibitem [{\citenamefont {Barbier}\ \emph {et~al.}(2010)\citenamefont {Barbier}, \citenamefont {Vasilopoulos},\ and\ \citenamefont {Peeters}}]{barbier2010kronig}%
  \BibitemOpen
  \bibfield  {author} {\bibinfo {author} {\bibfnamefont {M.}~\bibnamefont {Barbier}}, \bibinfo {author} {\bibfnamefont {P.}~\bibnamefont {Vasilopoulos}}, \ and\ \bibinfo {author} {\bibfnamefont {F.~M.}\ \bibnamefont {Peeters}},\ }\href {http://dx.doi.org/10.1103/PhysRevB.82.235408} {\bibfield  {journal} {\bibinfo  {journal} {Phys. Rev. B}\ }\textbf {\bibinfo {volume} {82}},\ \bibinfo {pages} {235408} (\bibinfo {year} {2010})}\BibitemShut {NoStop}%
\end{thebibliography}
\end{document}